\documentclass[preprint,prb,amsmath,amssymb,floatfix,nofootinbib]{revtex4}

\usepackage[colorlinks=true,urlcolor=black]{hyperref}

\hypersetup{backref = true,pagebackref = true,hyperindex = true, colorlinks = true,
  breaklinks = true, urlcolor = blue, linkcolor = blue, bookmarks = true,
  bookmarksopen = true, citecolor=red}

\usepackage{bm}
\usepackage{color}
\usepackage{dcolumn}
\usepackage{graphics}
\usepackage{graphicx}
\usepackage{subfigure}
\usepackage{slashed}
\usepackage{placeins}

\allowdisplaybreaks
\usepackage{pdfpages}
\usepackage{eso-pic}
\usepackage{everyshi}
\usepackage{pdflscape}

\newcommand{\bea}{\begin{eqnarray}}
\newcommand{\eea}{\end{eqnarray}}
\newcommand{\be}{\begin{equation}}
\newcommand{\ee}{\end{equation}}

\newcommand{\ar}{a_s}

\begin{document}

\title{Fractional Analytic QCD  beyond Leading Order}
       \author{A.\ V.~Kotikov$^{1}$ and I.A.~Zemlyakov$^{1,2}$
       }
       \affiliation{
$^1$Bogoliubov Laboratory of Theoretical Physics, Joint Institute for Nuclear Research, 141980 Dubna, Russia.\\
         $^2$Dubna State University,
Dubna, Moscow Region, Russia }

\date{\today}

\begin{abstract}
  Fractional analytic QCD is constructed beyond leading order
  using the standard inverse logarithmic expansion.
   It is shown that, contrary to the usual QCD coupling constant, for which this expansion can be used
   only for large values of its argument, in the case of analytic QCD, the inverse logarithmic expansion
   is applicable for all values of  the argument of the analytic coupling constant.
   We present four different views, two of which are based primarily on Polylogarithms and generalized Euler $\zeta$-functions, and the other two are based
   on dispersion integrals.
   The results obtained up to the 5th order of perturbation theory, have a compact form and do not contain complex special functions that were used to solve
   this problem earlier.
   As an example, we
   apply our results to study the polarized Bjorken sum rule, which is currently measured very accurately.

\end{abstract}

\maketitle

\section{ Introduction }

According to the general principles of (local) quantum field theory (QFT) \cite{Bogolyubov:1959bfo,Oehme:1994pv},
observables in the space-like domain can have
singularities only for negative values of their argument $Q^2$. On the other hand, for large values of $Q^2$, these observables are usually
represented as power expansions by the running coupling constant ({\it couplant}) $\alpha_s(Q^2)$, which, in turn, has a ghost singularity,
the so-called Landau pole, for $Q^2 = \Lambda^2$. To restore analyticity, this pole must be removed.

Indeed, the strong couplant $\alpha_s(Q^2)$ obeys  the renormalization group equation
\be
L\equiv \ln\frac{Q^2}{\Lambda^2} = \int^{\overline{a}_s(Q^2)} \, \frac{da}{\beta(a)},~~ \overline{a}_s(Q^2)=\frac{\alpha_s(Q^2)}{4\pi}\,
\label{RenGro}
\ee
with some boundary condition and the QCD $\beta$-function:
\be
\beta(\overline{a}_s) ~=~ -\sum_{i=0} \beta_i \overline{a}_s^{i+2} 
=-\beta_0  \overline{a}_s^{2} \, \Bigl(1+\sum_{i=1} b_i \beta_0^i \overline{a}_s^i\Bigr),~~ b_i=\frac{\beta_i}{\beta_0^{i+1}}\,, 
\label{beta}
\ee
where the first fifth coefficients, i.e. $\beta_i$ with $i\leq 4$, are exactly known \cite{Baikov:2016tgj,Herzog:2017ohr,Luthe:2017ttg}
(for convenience, they are listed in Appendix A).

Here we introduce a new definition of strong couplant:
\be
\ar(Q^2)=\frac{\beta_0\alpha_s(Q^2)}{4\pi}\,=\beta_0\,\overline{a}_s(Q^2)\,, 
\label{as}
\ee
where we add the first coefficient of the  QCD $\beta$-function to the $\ar$ definition, as is usually the case  in the case of  
of analytic couplants (see, e.g., Refs. \cite{ShS}-\cite{Cvetic:2008bn}).

So, already at leading order (LO), when $\ar(Q^2)=\ar^{(1)}(Q^2)$, we have from Eq. (\ref{RenGro})
\be
\ar^{(1)}(Q^2) = \frac{1}{L}\, ,
\label{asLO}
\ee
i.e. $\ar^{(1)}(Q^2)$ does contain a pole at $Q^2=\Lambda^2$.

In a series of papers \cite{ShS,MSS,Sh},
an effective approach
was developed
singularity without introducing extraneous IR regulators, such as
the effective gluon mass (see, e.g., \cite{Parisi:1979se,Cornwall:1981zr,GayDucati:1993fn,Mattingly:1992ud}).

The idea is based on the dispersion relation, which connects the new analytic couplant  $A_{\rm MA}(Q^2)$ with the spectral function
$r_{\rm pt}(s)$, obtained in the framework of perturbation theory. In LO, this gives the following
    \be
A^{(1)}_{\rm MA}(Q^2) 
= \frac{1}{\pi} \int_{0}^{+\infty} \, 
\frac{ d s }{(s + t)} \, r^{(1)}_{\rm pt}(s) \, ,
\label{disp_MA_LO}
\ee
where
\be
r^{(1)}_{\rm pt}(s)= {\rm Im} \; a_s^{(1)}(-s - i \epsilon) \,.
\label{SpeFun_LO}
\ee

So, let's repeat again: the spectral function is taken directly from perturbation theory, but the analytic  couplant $A_{\rm MA}(Q^2) $ is restored
using the dispersion relation (i.e. Eq. (\ref{disp_MA_LO}) at LO).
This approach is called {\it Minimal Approach} (MA) (see, e.g., \cite{Cvetic:2008bn})  or {\it Analytic Perturbation Theory} (APT) \cite{ShS,MSS,Sh}.
\footnote{An overview of other similar approaches can be found in \cite{Bakulev:2008td} including approaches \cite{Nesterenko:2003xb,Nesterenko:2004tg} close to APT.}


Thus, analytic QCD in its minimal version is a very convenient approach that combines the general (analytical) properties of quantum field quantities
and the results obtained within the framework of perturbative QCD, leading to the appearance of the MA couplant
$A_{\rm MA}(Q^2)$, close to the usual strong couplant $a_s(Q^2)$ in the limit of large values of its argument and completely different at $Q^2 \leq \Lambda^2$.

A further development of APT is the so-called fractional APT (FAPT), which extends the principles of constructing  to non-integer powers of couplant, which in
the QFT framework arise for many quantities having non-zero anomalous dimensions (see the famous papers \cite{BMS1,Bakulev:2006ex,Bakulev:2010gm},
some previous study \cite{Karanikas:2001cs} and
reviews in Ref. \cite{Bakulev:2008td}). This FAPT was developed mainly
for LO of perturbation theory, however, it was also used in higher orders by re-expanding the corresponding coupling constants in
terms of LO ones, as well as using some approximations. 

In this paper, we extend the FAPT to higher orders of perturbation theory using the so-called
$1/L$-expansion of the usual couplant. Note that for  an ordinary coupling constant, this expansion is applicable only for
large values of its argument $Q^2$, i.e. for $Q^2>>\Lambda^2$; however, in the case of an analytic coupling constant, the situation changes greatly and this
expansion is applicable for all values of the argument.  This is due to the fact that the non-leading expansion corrections disappear not only at $Q^2 \to \infty$,
but also at $Q^2 \to 0$,
\footnote{The absence of high-order corrections for $Q^2 \to 0$ was also discussed in Refs. \cite{ShS,MSS,Sh}.}
which leads to non-zero (small) corrections only in the domain of $Q^2 \sim \Lambda^2$ (see detailed discussions in Section 4 below).

Below we give four different representations for the MA couplant and its (fractional) derivatives in principle in any order of perturbation theory,
limiting ourselves to formulas of only the first five orders, all of whose parameters (related to the coefficients of the QCD $\beta$-function) are well known.
For applications, any of the proposed representations can be used, convenient in each specific case. 

The paper is organized as follows. In Section 2 we firstly review the basic properties of the usual strong couplant and its $1/L$-expansion. Section 3 contains
fractional derivatives (i.e. $\nu$-derivatives) of the usual strong couplant, which $1/L$-expansions can be represented as some operators acting on the
$\nu$-derivatives of the LO strong couplant. This is the key idea of this paper, which makes it possible to construct $1/L$-expansions of $\nu$-derivatives
of MA couplant for high-order perturbation theory, two different possibilities of which are presented in Section 4 and 5. In addition, Section 6 presents
two integral representations of the MA couplant at high orders of perturbation theory. 
One is based on the spectral density obtained
in high orders of perturbation theory, and the other is obtained using the above operators.
Sections 7 contains an application of this approach to the Bjorken sum rule. In conclusion, some final discussions are given. In addition, 
we have several Appendices.
For convenience, Appendix A lists the first five terms
\cite{Baikov:2016tgj,Herzog:2017ohr,Luthe:2017ttg} of the QCD $\beta$-function.
Appendix B contains formulas for reconstructing $\nu$-derivatives of the MA strong couplant in higher orders.
Appendices C and D present some alternative results for $\nu$-derivatives of the MA couplant, which may be useful for some applications.

\section{Strong coupling constant}
\label{strong}

As shown in Introduction, the strong couplant $a_s(Q^2)$ obeys the renormalization group equation (\ref{RenGro}).
When $Q^2>>\Lambda^2$, the Eq. (\ref{RenGro})
can be solved by iterations in the form of $1/L$-expansion
(we present the first five terms of the expansion in an agreement with the number of known coefficients $\beta_i$),
which can be represented in the following compact form
\be
a^{(1)}_{s,0}(Q^2) = \frac{1}{L_0},~~
a^{(i+1)}_{s,i}(Q^2) = 
a^{(1)}_{s,i}(Q^2) + \sum_{m=2}^i \, \delta^{(m)}_{s,i}(Q^2)
\,,~~(i=0,1,2,...)
\label{as}
\ee
where
\be
L_i=\ln
\frac{Q^2}{\Lambda_i^2}\,
\label{L}
\ee
and the symbol ``i'' shows the dependence of the parameter $\Lambda$ on the order of perturbation theory (see Eq. (\ref{Lambdas}) below).

The corrections $\delta^{(m)}_{s,k}(Q^2)$ can be represented as follows
\bea
&&\delta^{(2)}_{s,k}(Q^2) = - \frac{b_1\ln L_k}{L_k^2} ,~~
\delta^{(3)}_{s,k}(Q^2) =  \frac{1}{L_k^3} \, \Bigl[b_1^2(\ln^2 L_k-\ln L_k-1)+b_2\Bigr]\, , \nonumber \\
&&\delta^{(4)}_{s,k}(Q^2) =  \frac{1}{L_k^4} \, \left[b_1^3(-\ln^3 L_k+\frac{5}{2} \,\ln^2 L_k+2\ln L_k-\frac{1}{2})
   -3b_1b_2\ln L_k +\frac{b_3}{2}\right]\, , \nonumber \\
&&\delta^{(5)}_{s,k}(Q^2) = \frac{1}{L_k^5} \, \biggl[b_1^4(\ln^4 L_k-\frac{13}{3} \,\ln^3 L_k -\frac{3}{2} \,\ln^2 L_k+
    4\ln L_k+\frac{7}{6})
    \nonumber \\ &&  +3b^2_1b_2(2\ln^2 L_k-\ln L_k-1)
    - b_1b_3(2\ln L_k+\frac{1}{6} ) 
+\frac{1}{3}(b_4+5b_2^2)\biggr] \, .
\label{ds}
\eea

In Eqs. (\ref{as}) and (\ref{ds}) we show exactly that  at any order of perturbation theory, the couplant $\ar(Q^2)$ contains its own parameter $\Lambda$
of dimensional transmutation.
It relates with the normalization $\alpha_s(M_Z^2)$ as
\be
\Lambda_{i}=M_Z \, \exp\left\{-\frac{1}{2} \left[\frac{1}{a_s(M_Z^2)} + b_1\, \ln a_s(M_Z^2) +
\int^{\overline{a}_s(M_Z^2)}_0 \, da \, \left(\frac{1}{\beta(a)}+ \frac{1}{a^2(\beta_0+\beta_1 a)}\right)\right]\right\}\,.
\label{Lambdai}
\ee
where $M_Z=91.1876 \pm 0.0021$ GeV is the mass of $Z$-boson. We remind that
the value of $\alpha_s(Q^2)$ at some reference scale should be determined from experimental data. The current world
  average value for the coupling evaluated at the $Z$-boson mass scale, $\alpha_s(M_Z^2)$, as determined by the Particle Data Group (PDG),
  is $0.1176 \pm 0.0010$.
  \cite{PDG20}.

\subsection{$f$-dependence of the couplant $\ar(Q^2)$ }

We would like to note the coefficients $\beta_i$ depend on the number $f$ of active quarks, which changes
at thresholds $Q^2_f \sim m^2_f$, where some additional quark comes to play at $Q^2 > Q^2_f$. Here $m_f$ is the $\overline{MS}$ mass of $f$ quark, for example,
$m_b=4.18 +0.003-0.002$ GeV and $m_c=1.27 \pm 0.02$ GeV  from PDG20 \cite{PDG20}
\footnote{Strictly speaking, the quark masses in $\overline{MS}$-scheme are $Q^2$-dependent and $m_f=m_f(Q^2=m_f^2)$. The $Q^2$-dependence is quite slow and it is
  not shown in the present study.}
So, the coupling constant $a_s$ is $f$-dependent and the $f$-dependence can be taken into $\Lambda$, i.e. $\Lambda^f$ contribute to above Eqs.
(\ref{RenGro})
and (\ref{as}).
Moreover, Eq. (\ref{Lambdai}) can be used really for $\Lambda_{i}^{f=5}$, since at $Q^2=M_Z^2$ five quarks are active.

The relations between $\Lambda_{i}^{f}$ and $\Lambda_{i}^{f-1}$ can be obtained from Eq.(\ref{Lambdai}) with the replacement
$M_Z \to Q_f$ and the so-called decoupling relations, i.e. the relations between
$a_s(f,Q_f^2)$ and  $a_s(f-1,Q_f^2)$. In the $\overline{MS}$ scheme, the decoupling relations are known up to four-loop order
\cite{Chetyrkin:2005ia,Schroder:2005hy,Kniehl:2006bg} and they are usually used
at $Q_f^2=m_f^2$, where the relations are simplified (for a recent review, see e.g. \cite{FLAG,Enterria}).  

Here we will not consider the $f$-dependence of $\Lambda_{i}^{f}$ and $a_s(f,M_Z^2)$. This will be the subject of the next publication.
Since we will mainly consider the region of low $Q^2$, we will use the results for $\Lambda_{i}^{f=3}$, which we need to construct the analytic couplant
for small  $Q^2$ values.

\subsection{Discussions}

\begin{figure}[!htb]
\centering
\includegraphics[width=0.58\textwidth]{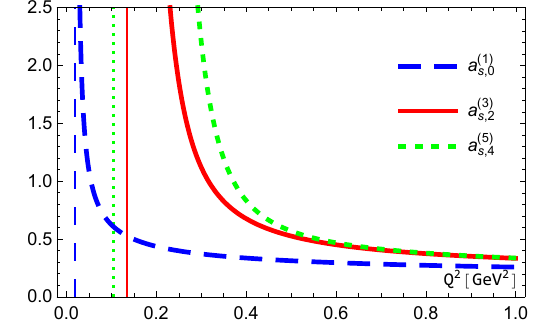}
\caption{\label{fig:as1352}
  The results for $a^{(i+1)}_{s,i}(Q^2)$ 
  and $(\Lambda_i^{f=3})^2$ (vertical lines)
  with $i=0,2,4$. Here and in the following figures,
  the $\Lambda_i^{f=3}$ values shown in (\ref{Lambdas}) are used.}
\end{figure}

In Fig. \ref{fig:as1352} one can see that the strong couplants  $a^{(i+1)}_{s,i}(Q^2)$ become to be singular at $Q^2=\Lambda_i^2$.
The
values of $\Lambda_0$ and $\Lambda_j$ $(j\geq 1)$ are very different
(see Eq. (\ref{Lambdas}) below): the values of $(\Lambda_i^{f=3})^2$  $(i=0,2,4)$ are also shown in Fig. 1 vertical lines.

We use the results for $\Lambda_i^{f=3}$ $(i=0,1,2,3)$ taken from the recent Ref. \cite{Chen:2021tjz}
\footnote{  The \cite{Chen:2021tjz} authors used the result of PDG20 $\alpha_s(M_Z)=0.1179(10)$. Now there is also the result PDG21 $\alpha_s(M_Z)=0.1179(9)$ which contains the
  same center value. Note that very close numerical relationships between $\Lambda_i$ were also obtained by \cite{Illa} for $\alpha_s(M_Z)=0.1168(19)$
  extracted by the ZEUS collaboration (see \cite{ZEUS}).}:
\be
\Lambda_0^{f=3}=142~~ \mbox{MeV},~~\Lambda_1^{f=3}=367~~ \mbox{MeV},~~\Lambda_2^{f=3}=324~~ \mbox{MeV},~~\Lambda_3^{f=3}=328~~ \mbox{MeV}\,.
\label{Lambdas}
\ee
We use also $\Lambda_4=\Lambda_3$, since in highest orders $\Lambda_i$ values become very similar.

\section{Fractional derivatives}

Following \cite{Cvetic:2006mk,Cvetic:2006gc},
we introduce the derivatives (in the $(i+1)$-order of perturbation theory)
\be
\tilde{a}^{(i+1)}_{n+1}(Q^2)=\frac{(-1)^n}{
  n!} \, \frac{d^n a^{(i+1)}_s(Q^2)}{(dL)^n} \, ,
\label{tan+1}
\ee
which will be very convenient in the case of the analytical QCD
(see e.g. Ref. \cite{Kotikov:2022swl} and discussions therein).

The series of derivatives $\tilde{a}_{n}(Q^2)$ can successfully replace the corresponding series of the $\ar$-powers. Indeed, every
derivative decrease the power of $\ar$ but it comes together with the additional $\beta$-function $\sim \ar^2$, appeared during the derivative.
So, every application of derivative produces the additional $\ar$,
and, thus, indeed the series of derivatives can be used instead of the series of the $\ar$-powers.

At LO, the series of derivatives $\tilde{a}_{n}(Q^2)$ exactly coincide with $\ar^{n}$. Beyond LO, 
the relation between $\tilde{a}_{n}(Q^2)$ and $\ar^{n}$ was established in Ref. \cite{Cvetic:2006gc,Cvetic:2010di} and extended to the fractional case,
where $n \to$ a non-integer $\nu $, in Ref. \cite{GCAK}.

Now we consider the
$1/L$ expansion of $\tilde{a}^{(k)}_{\nu}(Q^2)$, which can be done in two different ways:\\
1. We can differentiate $n$ times the above results (\ref{as}) and (\ref{ds}) and later to transform the obtained results to the non-integer
values $\nu$ in an agreement with Ref. \cite{GCAK}.\\
2. We can firstly to find the $\nu$-powers of the results (\ref{as}) and (\ref{ds}) and later to reconstruct $\tilde{a}^{(k)}_{\nu}(Q^2)$ using
the relations between $\tilde{a}_{\nu}$ and $\ar^{\nu}$ obtained in \cite{GCAK}.

We use the second possibility. The evaluation is considered in details in Appendix B. Here we present only the final results of calculations,
which have the following form
\footnote{  The extension (\ref{tdmp1N}) is very similar to those used in Refs. \cite{BMS1,Bakulev:2006ex} for the expansion of
  ${\bigl({a}^{(i+1)} _{s,i}(Q^2)\bigr)}^ {\nu}$ in terms of powers of $a^{(1)}_{s,i}(Q^2)$. }
:
\bea
&&\tilde{a}^{(1)}_{\nu,0}(Q^2)={\bigl(a^{(1)}_{s,0}(Q^2)\bigr)}^{\nu} = \frac{1}{L_0^{\nu}},~~
\tilde{a}^{(i+1)}_{\nu,i}(Q^2)=\tilde{a}^{(1)}_{\nu,i}(Q^2) + \sum_{m=1}^{i}\, C_m^{\nu+m}\, \tilde{\delta}^{(m+1)}_{\nu,i}(Q^2),~~\nonumber\\
&&\tilde{\delta}^{(m+1)}_{\nu,i}(Q^2)=
\hat{R}_m \, \frac{1}{L_i^{\nu+m}},~~C_m^{\nu+m}=\frac{\Gamma(\nu+m)}{m!\Gamma(\nu)}\,,
\label{tdmp1N}
\eea
where
\bea
&&\hat{R}_1=b_1 \Bigl[\hat{Z}_1(\nu)+ \frac{d}{d\nu}\Bigr],~~
\hat{R}_2=b_2 + b_1^2 \Bigl[\frac{d^2}{(d\nu)^2} +2 \hat{Z}_1(\nu+1)\frac{d}{d\nu} + \hat{Z}_2(\nu+1 )\Bigr], \nonumber\\
&&\hat{R}_3=\frac{b_3}{2} + 3b_2b_1\Bigl[Z_1(\nu+2)-\frac{11}{6}+\frac{d}{d\nu}\Bigr]\nonumber \\
&&+ b_1^3 \Bigl[ \frac{d^3}{(d\nu)^3}+3\hat{Z}_1(\nu+2) \frac{d^2}{(d\nu)^2} +3 \hat{Z}_2(\nu+2)\frac{d}{d\nu} + \hat{Z}_3(\nu+2 )\Bigr], \nonumber\\
&&\hat{R}_4=\frac{1}{3}\,\bigl(b_4+5b_2^2\bigr) + 2b_3b_1\Bigl[Z_1(\nu+3)-\frac{13}{6}+\frac{d}{d\nu}\Bigr]\nonumber \\
&&+ 6b_1^2b_2 \Bigl[ \frac{d^2}{(d\nu)^2}+2\left(Z_1(\nu+3)-\frac{11}{6}\right) \frac{d}{d\nu} + Z_2(\nu+3) -\frac{11}{3} \, Z_1(\nu+3)+\frac{38}{9}
    \Bigr]\nonumber\\
&&+ b_1^4 \Bigl[\frac{d^4}{(d\nu)^4}+4\hat{Z}_1(\nu+3) \frac{d^3}{(d\nu)^3} +6\hat{Z}_2(\nu+3) \frac{d^2}{(d\nu)^2}
  +4 \hat{Z}_3(\nu+3)\frac{d}{d\nu} + \hat{Z}_4(\nu+3 )\Bigr]\, .
\label{hR_i}
\eea

The representation (\ref{tdmp1N}) of the $\tilde{\delta}^{(m+1)}_{\nu,i}(Q^2)$ corrections as $\hat{R} _m$-operators is very important to use.
This will make it possible to present high-order results for the ($1/L$-expansion of) the  analytic couplant in a similar way.

We would like to note that,
using quite complicated forms for the powers of couplant ${[a_s^{(i+1)}(Q^2)]}^{\nu}$ and for the coefficients $k_m(\nu)$, shown in Appendix B,
we have got a rather compact form
for the derivatives $\tilde{a}^{(i+1)}_{\nu}(Q^2)$.


\section{Minimal analytic coupling}

There are several ways to obtain analytical versions of the strong couplant $a_s$ (see, e.g. \cite{Bakulev:2008td}). Here we will follow
MA approach \cite{ShS,MSS,Sh} as discussed in Introduction.
To the fractional case,  the MA approach was generalized by Bakulev, Mikhailov and Stefanis (hereinafter referred to as the BMS approach), which was
presented
in three famous papers \cite{BMS1,Bakulev:2006ex,Bakulev:2010gm}
(see also a previous paper \cite{Karanikas:2001cs}, the reviews \cite{Bakulev:2008td,Cvetic:2008bn} and Mathematica package in \cite{Bakulev:2012sm}). 

We first show the leading order BMS results, and later we will go beyond LO, following our results for the usual strong couplant obtained in the previous section
(see Eq. (\ref{tdmp1N})).

\subsection{LO}

\begin{figure}[!htb]
\centering
\includegraphics[width=0.58\textwidth]{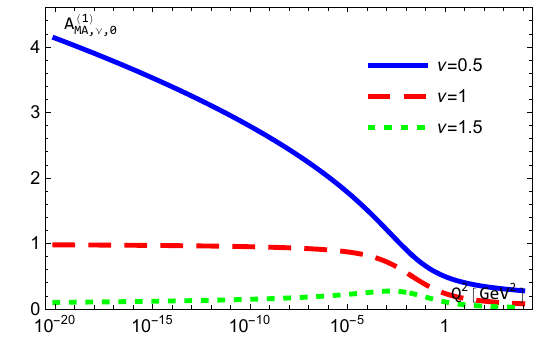}
    \caption{\label{fig:A1}
      The results for $A^{(1)}_{\rm MA,\nu,0}(Q^2)$ with $\nu=0.5,1,1.5$.
    }
\end{figure}

The LO minimal analytic coupling $A^{(1)}_{{\rm MA},\nu}$
have the form  \cite{BMS1}
\be
A^{(1)}_{{\rm MA},\nu,0}(Q^2) = {\left( a^{(1)}_{\nu,0}(Q^2)\right)}^{\nu} - \frac{{\rm Li}_{1-\nu}(z_0)}{\Gamma(\nu)}=
\frac{1}{L_0^{\nu}}- \frac{{\rm Li}_{1-\nu}(z_0)}{\Gamma(\nu)} \equiv \frac{1}{L_0^{\nu}}-\Delta^{(1)}_{\nu,0}\,,~~ z_k=\frac{\Lambda_k}{Q^2}\,,
\label{tAMAnu}
\ee
where
\be
   {\rm Li}_{\nu}(z)=\sum_{m=1}^{\infty} \, \frac{z^m}{m^{\nu}}=  \frac{z}{\Gamma(\nu)} \int_0^{\infty} 
\frac{ dt \; t^{\nu -1} }{(e^t - z)}
   \label{Linu}
\ee
is the Polylogarithmic function. For the cases $\nu=0.5,1,1.5$, $A^{(1)}_{{\rm MA},\nu,0}(Q^2)$ is shown in Fig. \ref{fig:A1}.
\footnote{
  Strictly speaking, the value of the parameter $\Lambda$ is obtained by fitting experimental data.
    To obtain its values within the framework of analytical QCD, it is necessary to fit experimental data for various processes using,
    for example, formulas obtained in this paper that simplify the form of higher-order terms. This, however, requires additional special
    research. Therefore, in this article we use the value $\Lambda_{f=3}$ obtained in the framework of a conventional perturbative QCD.}
%
It is clearly seen that $A^{(1)}_{{\rm MA},\nu,0}(Q^2\to 0)$ agree with its asymptotic values:
\be
A^{(1)}_{{\rm MA},\nu,0}(Q^2= 0) = \left\{
\begin{array}{c}
0 ~~\mbox{when}~~ \nu >1, \\
1 ~~\mbox{when}~~ \nu =1, \\
\infty ~~\mbox{when}~~ \nu <1, 
\end{array}
\right.
   \label{AQ=0}
\ee
obtained in Ref. \cite{Ayala:2018ifo}.

For $\nu=1$ we recover the famous Shirkov-Solovtsov result  \cite{ShS,Sh}:
\be
A^{(1)}_{\rm MA,0}(Q^2) \equiv A^{(1)}_{\rm MA,\nu=1,0}(Q^2) =  a^{(1)}_{s,0}(Q^2) - \frac{z_0}{1-z_0}=\frac{1}{L_0}- \frac{z_0}{1-z_0}\, ,
\label{tAM1}
\ee
since
\be
   {\rm Li}_{0}(z)= \frac{z}{1-z} \, .
\label{Li0.1}
\ee
Note that the result (\ref{tAM1}) can be taken directly for the integral form (\ref{disp_MA_LO}), as it was in Ref. \cite{ShS,Sh}.

\subsection{Beyond LO}

Following to the
representation (\ref{tAMAnu}) for the LO analytic couplant, 
we consider the difference between the derivatives of usual and (minimal) analytic couplants, shown in Eq.(\ref{tan+1}) and in
\be
\tilde{A}_{{\rm MA},n+1}(Q^2)=\frac{(-1)^n}{
  n!} \, \frac{d^n A_{\rm MA}(Q^2)}{(dL)^n} \, ,
\label{tanMA+1}
\ee
respectively.

Using the results (\ref{tdmp1N}) by analogy with the usual couplant we have
for the differences of fractional derivatives of usual and analytic couplants
\be
\tilde{\Delta}^{(i+1)}_{\nu,i} \equiv \tilde{a}^{(i+1)}_{\nu,i} - \tilde{A}^{(i+1)}_{{\rm MA},\nu,i}
\label{tDeltaMAnu}
\ee
the following results
\be
\tilde{\Delta}^{(i+1)}_{\nu,i}
=\tilde{\Delta}^{(1)}_{\nu,i}
+\sum_{m=1}^i C_m^{\nu+m} \, \hat{R}_m \left( \frac{{\rm Li}_{-\nu-m+1}(z_i)}{\Gamma(\nu+m)}\right)
\, ,
\label{tAMAnu.1}
\ee
where the operators $\hat{R}_i$ $(i=1,2,3,4)$ are shown above in Eq. (\ref{hR_i}).
The relations (\ref{tAMAnu.1})
reflect the fact that the MA procedure (\ref{tAMAnu}) and the operation $d/(d\nu)$ commute.

Thus, to obtain (\ref{tAMAnu.1}) we propose that the form (\ref{tdmp1N})  for the usual couplant $a_s$ at high orders
is exactly applicable (exactly in the same way) also to the case of the (MA) couplant.


After some evaluations, we obtain the following expressions without operators
\be
\tilde{\Delta}^{(i+1)}_{\nu,i}
=\tilde{\Delta}^{(1)}_{\nu,i}
+\sum_{m=1}^i C_m^{\nu+m} \, \overline{R}_m(z_i) \left( \frac{{\rm Li}_{-\nu-m+1}(z_i)}{\Gamma(\nu+m)}\right)
\, ,
\label{tAMAnu.2}
\ee
where
\bea
&&\overline{R}_1(z)=b_1\Bigl[\gamma_{\rm E}-1+{\rm M}_{-\nu,1}(z)\Bigr], \nonumber \\
&&\overline{R}_2(z)=b_2 + b_1^2\Bigl[{\rm M}_{-\nu-1,2}(z) + 2(\gamma_{\rm E}-1){\rm M}_{-\nu-1,1}(z) +  (\gamma_{\rm E}-1)^2-
    \zeta_2\Bigr], \nonumber \\
&&\overline{R}_3(z)=\frac{b_3}{2} +3b_2b_1\left[ \gamma_{\rm E}-\frac{11}{6}+{\rm M}_{-\nu-2,1}(z)\right] + b_1^3\biggl[{\rm M}_{-\nu-2,3}(z) + 3(\gamma_{\rm E}-1) \, {\rm M}_{-\nu-2,2}(z) \nonumber \\
&&\hspace{2cm} +  3\Bigl((\gamma_{\rm E}-1)^2-
  \zeta_2\Bigr)  \, {\rm M}_{-\nu-2,1}(z) + (\gamma_{\rm E}-1)\Bigl((\gamma_{\rm E}-1)^2-3\zeta_2\Bigr)+2 \zeta_3 \biggr]
, \nonumber \\
&&\overline{R}_4(z)=\frac{1}{3}\left(b_4+5b_2^2\right) +2b_3b_1\left[ \gamma_{\rm E}-\frac{13}{6}+{\rm M}_{-\nu-3,1}(z)\right]
\nonumber \\&&
+6b_2b_1^2\left[ \gamma^2_{\rm E}-\frac{11}{3}\gamma_{\rm E}-\zeta_2 +\frac{38}{9}
  + 2\left(\gamma_{\rm E}-\frac{11}{6}\right){\rm M}_{-\nu-3,1}(z)+{\rm M}_{-\nu-3,2}(z)\right]
\nonumber \\&&
+ b_1^4\biggl[{\rm M}_{-\nu-3,4}(z) + 4(\gamma_{\rm E}-1) \, {\rm M}_{-\nu-3,3}(z) +  6\Bigl((\gamma_{\rm E}-1)^2-   \zeta_2\Bigr)  \, {\rm M}_{-\nu-3,2}(z)  \nonumber \\
  &&\hspace{2cm}
  + 4\Bigl((\gamma_{\rm E}-1)\Bigl((\gamma_{\rm E}-1)^2-3\zeta_2\Bigr)+2 \zeta_3\Bigr)
  \, {\rm M}_{-\nu-3,1}(z) \nonumber \\
  &&\hspace{2cm}+ (\gamma_{\rm E}-1)^2\Bigl((\gamma_{\rm E}-1)^2-6\zeta_2\Bigr)+8(\gamma_{\rm E}-1) \zeta_3 + 3\zeta_2^2-6\zeta_4  \biggr]
\label{oRi}
\eea
and
\be
   {\rm Li}_{\nu,k}(z)=(-1)^k\,\frac{d^k}{(d\nu)^k}  \,{\rm Li}_{\nu}(z) =
   \sum_{m=1}^{\infty} \, \frac{z^m\ln^k m}{m^{\nu}},~~{\rm M}_{\nu,k}(z)=\frac{{\rm Li}_{\nu,k}(z)}{{\rm Li}_{\nu}(z)} \, .
   \label{Mnuk}
\ee
We see that the $\Psi(\nu)$-function and its derivatives have completely canceled out. 
Note that another form for $\tilde{\Delta}^{(m+1)}_{\nu,i}(Q^2)$ is given in Appendix C.

So, we have for MA analytic couplants $\tilde{A}^{(i+1)}_{{\rm MA},\nu}$ the following expressions:
\be
\tilde{A}^{(i+1)}_{{\rm MA},\nu,i}(Q^2) = \tilde{A}^{(1)}_{{\rm MA},\nu,i}(Q^2) + \sum_{m=1}^{i}  \, C^{\nu+m}_m \tilde{\delta}^{(m+1)}_{{\rm ma},\nu,i}(Q^2) \,
\label{tAiman}
\ee
where
\bea
&&\tilde{A}^{(1)}_{{\rm MA},\nu,i}(Q^2) = \tilde{a}^{(1)}_{\nu,i}(Q^2) -  \frac{{\rm Li}_{1-\nu}(z_i)}{\Gamma(\nu)},
~~\nonumber \\
&&\tilde{\delta}^{(m+1)}_{{\rm MA},\nu,i}(Q^2)= \tilde{\delta}^{(m+1)}_{\nu,i}(Q^2) -  \overline{R}_m(z_i)   \, \frac{{\rm Li}_{-\nu+1-m}(z_i)}{\Gamma(\nu+m)}
\label{tdAman}
\eea
and $\tilde{\delta}^{(k+1)}_{\nu,m}(Q^2)$
are given in Eq. (\ref{tdmp1N}).

\subsection{The case $\nu=1$
}

For the case $\nu=1$, at LO we have Eq. (\ref{tAM1}) and above LO we can apply above results (\ref{tAiman}) - (\ref{tdAman})
to the case $\nu=1$:
\be
A^{(i+1)}_{{\rm MA},i}(Q^2)\equiv \tilde{A}^{(i+1)}_{{\rm MA},\nu=1,i}(Q^2) = A^{(1)}_{{\rm MA},i}(Q^2) + \sum_{m=1}^{i}  \, \tilde{\delta}^{(m+1)}_{{\rm MA},1,i}(Q^2) \,
\label{tAiman.1}
\ee
where (according to (\ref{tAM1}))
\bea
&&A^{(1)}_{{\rm MA},i}(Q^2) = \tilde{a}^{(1)}_{\nu=1,i}(Q^2) -  {\rm Li}_{0}(z_i)= a^{(1)}_{s,i}(Q^2) -  {\rm Li}_{0}(z_i),
~~\nonumber \\
&&\tilde{\delta}^{(m+1)}_{{\rm MA},1,i}(Q^2)= \tilde{\delta}^{(m+1)}_{1,i}(Q^2) -  \overline{R}_m(z_i)   \, \frac{{\rm Li}_{-m}(z_i)}{m!}
= \tilde{\delta}^{(m+1)}_{1,i}(Q^2) - \frac{P_{m,1}(z_i)}{m!}
\label{tdAmanA}
\eea
and $P_{m,1}(z_i)$ are given in Eqs. (\ref{PkZm}) and (\ref{Pkz}) at $\nu=1$ and also
\bea
   &&{\rm Li}_{-1}(z)= \frac{z}{(1-z)^2},~~{\rm Li}_{-2}(z)= \frac{z(1+z)}{(1-z)^3},~~{\rm Li}_{-3}(z)= \frac{z(1+4z+z^2)}{(1-z)^4},~~\nonumber \\
&&
   {\rm Li}_{-4}(z)= \frac{z(1+z)(1+10z+z^2)}{(1-z)^5}\, .
\label{Lii.1}
\eea

The results (\ref{tdAmanA}) and (\ref{Lii.1})
can be used for phenomenological studies beyond LO in the framework of the minimal analytic QCD.

\subsection{Discussions}

\begin{figure}[!htb]
\centering
\includegraphics[width=0.58\textwidth]{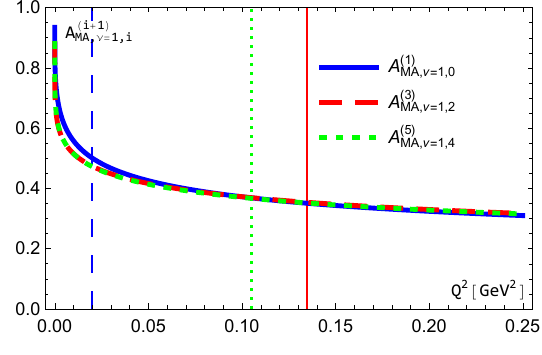}
\caption{\label{fig:A123}
  The results for $A^{(i+1)}_{\rm MA,\nu=1,i}(Q^2)$
  and $(\Lambda_i^{f=3})^2$ (vertical lines)
  with $i=0,2,4$.}
\end{figure}

\begin{figure}[!htb]
\centering
\includegraphics[width=0.58\textwidth]{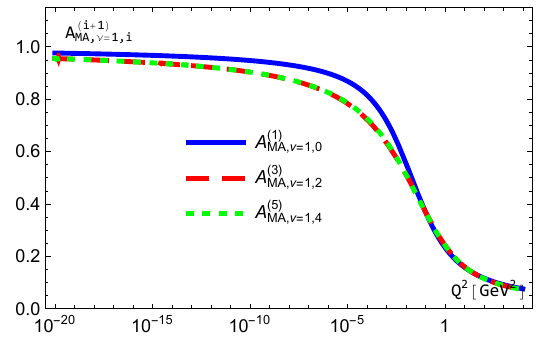}
    \caption{\label{fig:A123LOG}
      The results for $A^{(i+1)}_{\rm MA,\nu=1,i}(Q^2)$ ($i=0,1,2$)  but with the logarithmic scale.
    }
\end{figure}

\begin{figure}[!htb]
\centering
\includegraphics[width=0.58\textwidth]{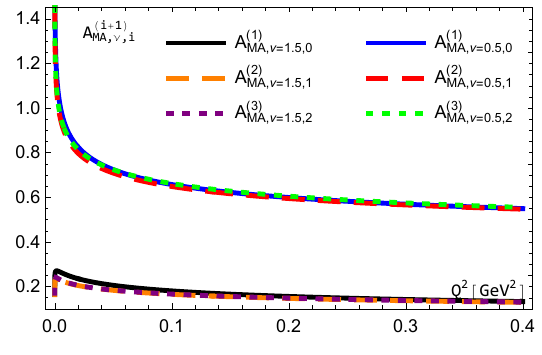}
    \caption{\label{fig:fracALL}
      The results for $A^{(i+1)}_{\rm MA,\nu=0.5,i}(Q^2)$ and $A^{(i+1)}_{\rm MA,\nu=1.5,i}(Q^2)$ with $i=0,1,2$.
    }
\end{figure}

\begin{figure}[!htb]
\centering
\includegraphics[width=0.58\textwidth]{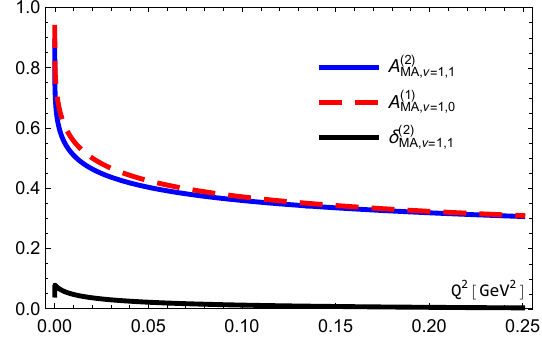}
    \caption{\label{fig:Adelta2}
      The results for $A^{(1)}_{\rm MA,\nu=1,0}(Q^2)$, $A^{(2)}_{\rm MA,\nu=1,1}(Q^2)$ and $\delta^{(2)}_{\rm MA,\nu=1,1}(Q^2)$.
    }
\end{figure}

\begin{figure}[!htb]
\centering
\includegraphics[width=0.58\textwidth]{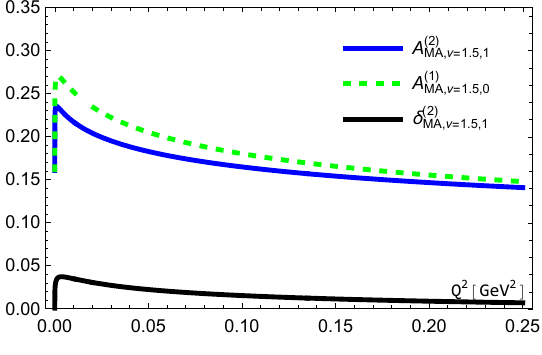}
    \caption{\label{fig:Afrac2}
      The results for $A^{(1)}_{\rm MA,\nu=1.5,0}(Q^2)$, $A^{(2)}_{\rm MA,\nu=1.5,1}(Q^2)$ and $\delta^{(2)}_{\rm MA,\nu=1.5,1}(Q^2)$.
    }
\end{figure}

\begin{figure}[!htb]
\centering
\includegraphics[width=0.58\textwidth]{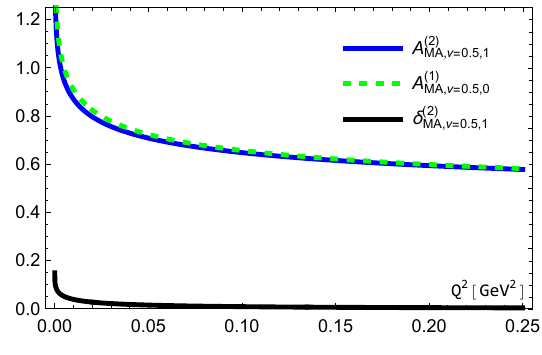}
    \caption{\label{fig:Afrac3}
      The results for $A^{(1)}_{\rm MA,\nu=0.5,0}(Q^2)$, $A^{(2)}_{\rm MA,\nu=0.5,1}(Q^2)$ and $\delta^{(2)}_{\rm MA,\nu=0.5,1}(Q^2)$.
    }
\end{figure}

\begin{figure}[!htb]
\centering
\includegraphics[width=0.58\textwidth]{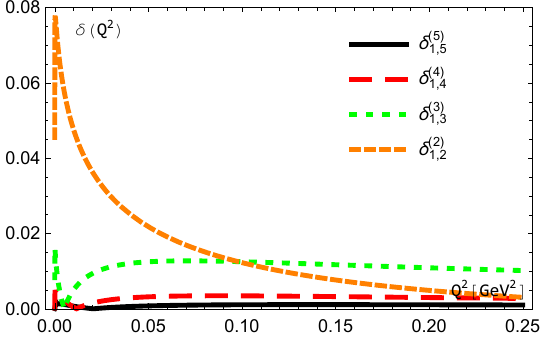}
    \caption{\label{fig:deltaALL}
      The results for $\delta^{(i+1)}_{\rm MA,\nu=1,i}(Q^2)$ with $i=1,2,3,4$.}
\end{figure}

\begin{figure}[!htb]
\centering
\includegraphics[width=0.58\textwidth]{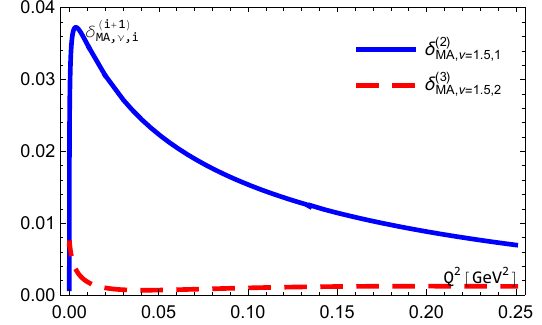}
    \caption{\label{fig:deltafrac32sec}
      The results for $\delta^{(i+1)}_{\rm MA,\nu=1.5,i}(Q^2)$ with $i=1,2$.}
\end{figure}

From Figs. \ref{fig:A123} and \ref{fig:A123LOG} we can see differences between $A^{(i+1)}_{\rm MA,\nu=1,i}(Q^2)$ with $i=0,2,4$,
which are rather small and have nonzero values around the
position $Q^2=\Lambda_i^2$. Similar situation exists also at the cases  $\nu=0.5$ and $\nu=1.5$
(see Fig. \ref{fig:fracALL}).
In Fig. \ref{fig:A123} the values of $(\Lambda_i^{f=3})^2$ $(i=0,2,4)$ are shown by vertical lines (as in Fig. 1).

Figs. \ref{fig:Adelta2}, \ref{fig:Afrac2} and \ref{fig:Afrac3} show the results for  $A^{(1)}_{\rm MA,\nu,0}(Q^2)$ and $A^{(2)}_{\rm MA,\nu,1}(Q^2)$ and their difference
$\delta^{(2)}_{\rm MA,\nu,1}(Q^2)$, which is essentially less then the couplants theirselves. This is shown for three different $\nu$-values: $0.5, 1, 1,5$.
From Figs. \ref{fig:fracALL}, \ref{fig:Afrac2} and \ref{fig:Afrac3} it is clear that for $Q^2\to 0$ the asymptotic behaviors of $A^{(1)}_{\rm MA,\nu, 0}(Q^2)$, 
$A^{(2)}_{\rm MA,\nu,1}(Q^2)$  and $A^{(3)}_{\rm MA,\nu,2}(Q^2)$ coincide (and are equal to those considered in (\ref{AQ=0})), i.e.
the differences $\delta^{(2)}_{\rm MA,\nu=1,1}(Q^2\to 0)$ and $\delta^{(3)}_{\rm MA,\nu=1,2}(Q^2\to 0)$ are negligible. 
Also Fig. \ref{fig:deltaALL} shows the differences $\delta^{(i+1)}_{\rm MA,\nu=1,i}(Q^2)$
$(i\geq 2)$ essentially less then $\delta^{(2)}_{\rm MA,\nu=1,1}(Q^2)$. From Fig. \ref{fig:deltafrac32sec} we can see a similar property for $\delta^{(i+1)}_{\rm MA,\nu=1.5,i}(Q^2)$
and $\delta^{(2)}_{\rm MA,\nu=1.5,1}(Q^2)$.

Thus, we can conclude that contrary to the case of the usual couplant, considered in Fig. 1,
the $1/L$-expansion of the MA couplant is very good approximation at any $Q^2$ values.
Moreover, the differences between $A^{(i+1)}_{\rm MA,\nu=1,i}(Q^2)$ and $A^{(1)}_{\rm MA,\nu=1,0}(Q^2)$ are small.
So, the expansions of  $A^{(i+1)}_{\rm MA,\nu=1,i}(Q^2)$ $i\geq 1$ through the one $A^{(1)}_{\rm MA,\nu=1,0}(Q^2)$ done in Refs. \cite{BMS1,Bakulev:2006ex,Bakulev:2010gm} are
very good approximations.
Also
the approximation
\be
A^{(i+1)}_{\rm MA,\nu=1,i}(Q^2)=A^{(1)}_{\rm MA,\nu=1,0}(k_iQ^2),~~(i=1,2)\,,
\label{Appro}
\ee
introduced in \cite{Pasechnik:2009yc,Khandramai:2011zd} and used in \cite{Kotikov:2010bm} is very convenient, too.
Indeed, since the corrections $\delta^{(i+1)}_{\rm MA,\nu=1,i}(Q^2)$ are very small, then for example from Eq. (\ref{tdAmanA}) one can see that the MA couplants
$A^{(i+1)}_{\rm MA,\nu=1,i}(Q^2)$ are very similar to the LO ones taken with the corresponding $\Lambda_i$.

\section{MA
  coupling:
  the form is convenient for $Q^2 \sim \Lambda_k^2$
  .}

The results (\ref{tAMAnu.2}) for analytic couplant is very convenient at the range at large and at small $Q^2$ values. For $Q^2 \sim \Lambda_i^2$
the both parts, the standard strong couplant and the additional term,
have singularities, which are cancelled in its sum. So, some numerical applications of the results
(\ref{tAMAnu.2})
can be complicated. So, here we present another form, which is very useful at $Q^2 \sim \Lambda_i^2$ and can be used also for any $Q^2$ values, excepting
the ranges of very large and very small $Q^2$ values. As in the previous section, we will present firstly LO results taken from \cite{BMS1} and later to
extend them beyond LO.

\subsection{LO}

The LO minimal analytic coupling $A^{(1)}_{{\rm MA},\nu}(Q^2)$ \cite{ShS,MSS,Sh}
have also the another form  \cite{BMS1}
\be
A^{(1)}_{{\rm MA},\nu}(Q^2) =
\frac{(-1)}{\Gamma(\nu)} \, \sum_{r=0}^{\infty} \zeta(1-\nu-r) \, \frac{(-L)^r}{r!}~~~ (L<2\pi),
\label{tAMAnuNew}
\ee
where Euler functions $\zeta(\nu)$ are
\be
\zeta(\nu)
     =\sum_{m=1}^{\infty} \, \frac{1}{m^{\nu}}={\rm Li}_{\nu}(z=1)
   \label{ze_nu}
\ee

The result  (\ref{tAMAnuNew}) has been obtained in Ref. \cite{BMS1} considering the property of the Lerch function, which can be considered as
a generalization of Polylogarithms (\ref{Linu}).
The form (\ref{tAMAnuNew}) is very convenient at low $L$ values, t.e. at $Q^2 \sim \Lambda^2$. Moreover, we can use the relation between
$\zeta(1-\nu-r)$ and $\zeta(\nu+r)$ functions
\be
\zeta(1-\nu-r)= \frac{2\Gamma(\nu+r)}{(2\pi)^{\nu+r}}\, Sin\left[\frac{\pi}{2}(1-\nu-r)\right] \, \zeta(\nu+r)
  \label{ze_nu.1}
\ee

For $\nu=1$ we have
\be
A^{(1)}_{\rm MA}(L) = - \, \sum_{r=0}^{\infty} \zeta(-r) \, \frac{(-L)^r}{r!}
\label{tAMA1New}
\ee
with
\be
 \zeta(-r) = (-1)^{r} \, \frac{B_{r+1}}{r+1}
\label{ze_r}
\ee
and $B_{r+1}$ are Bernoulli numbers.

Using the properties of Bernoulli numbers ($\delta^0_m$ is Kronecker symbol), we have for even $r=2m$ and for odd  $r=1+2l$ values
\be
\zeta(-2m) = -\frac{\delta^0_m}{2},~~\zeta(-(1+2l))= - \frac{B_{2(l+1)}}{2(l+1)} \,.
\label{ze_r1}
\ee
Thus, we have for $A^{(1)}_{\rm MA}(Q^2)$ the following results
\be
A^{(1)}_{\rm MA}(Q^2) = \frac{1}{2} \, \left(1
+ \, \sum_{l=0}^{\infty} \frac{B_{2(l+1)}}{l+1} \, \frac{(-L)^{2l+1}}{(2l+1)!}\right)=\frac{1}{2} \, \left(1
+ \, \sum_{s=1}^{\infty} \frac{B_{2s}}{s} \, \frac{(-L)^{2s-1}}{(2s-1)!}\right) \, ,
\label{tAMA1New.1}
\ee
with $s=l+1$.

\subsection{Beyond LO}

Now we consider the derivatives of (minimal) analytic couplants $\tilde{A}^{(1)}_{{\rm MA},\nu}$, shown in Eq.(\ref{tanMA+1}), as in Eq. (\ref{tAiman}), i.e.
\be
\tilde{A}^{(i+1)}_{{\rm MA},\nu,i}(Q^2) = \tilde{A}^{(1)}_{{\rm MA},\nu,i}(Q^2) + \sum_{m=1}^{i}  \, C^{\nu+m}_m \tilde{\delta}^{(m+1)}_{{\rm MA},\nu,i}(Q^2) \,
\label{tAimanNew}
\ee
where $\tilde{A}^{(1)}_{{\rm MA},\nu,i}=A^{(1)}_{{\rm MA},\nu}$ is given above in (\ref{tAMAnuNew}) with $L \to L_{i}$ and
\be
\tilde{\delta}^{(m+1)}_{{\rm MA},\nu,i}(Q^2)= \hat{R}_m   \, A^{(1)}_{{\rm MA},\nu+m,i} \, ,
\label{tdAmanNew}
\ee
where operators $\hat{R}_m$ are given above in (\ref{hR_i}).

After come calculations we have
\be
\tilde{\delta}^{(m+1)}_{{\rm MA},\nu,k}(Q^2)=
\frac{(-1)}{\Gamma(\nu+m)} \, \sum_{r=0}^{\infty} \tilde{R}_m(\nu+r) \, \frac{(-L_k)^r}{r!}
\label{tdAmanNew}
\ee
where (in an agreement with (\ref{oRi}) and (\ref{Pkz}))
\bea
&&\tilde{R}_1(\nu+r)=b_1\Bigl[(\gamma_{\rm E}-1)\zeta(-\nu-r)+\zeta_1(-\nu-r)\Bigr], \nonumber \\
&&\tilde{R}_2(\nu+r)=b_2\zeta(-\nu-r-1) + b_1^2\Bigl[\zeta_2(-\nu-r-1) + 2(\gamma_{\rm E}-1)\zeta_1(-\nu-r-1) \nonumber \\
  &&\hspace{0.5cm} +  \bigl[(\gamma_{\rm E}-1)^2-
    \zeta_2\bigr]\zeta(-\nu-r-1)\Bigr], \nonumber \\
            &&\tilde{R}_3(\nu+r)=\frac{b_3}{2}\zeta(-\nu-r-2) +3b_2b_1\left[ \bigl(\gamma_{\rm E}-\frac{11}{6}\bigr)\zeta(-\nu-r-2)+\zeta_1(-\nu-r-2)\right]
\nonumber\\  &&\hspace{0.5cm}   + b_1^3\biggl[\zeta_3(-\nu-r-2) + 3(\gamma_{\rm E}-1) \, \zeta_2(-\nu-r-2)
 +  3\Bigl[(\gamma_{\rm E}-1)^2-
   \zeta_2\Bigr]  \, \zeta_1(-\nu-r-2)
 \nonumber\\  &&\hspace{0.5cm} + \bigl[(\gamma_{\rm E}-1)\Bigl((\gamma_{\rm E}-1)^2-3\zeta_2\Bigr)+2 \zeta_3\bigr]\, \zeta(-\nu-r-2) \biggr]
, \nonumber \\
&&\tilde{R}_4(\nu+r)=\frac{1}{3}\left(b_4+5b_2^2\right)\zeta(-\nu-r-3)
+2b_3b_1\biggl[ \bigl(\gamma_{\rm E}-\frac{13}{6}\bigr)\zeta(-\nu-r-3)
   \nonumber\\  &&\hspace{0.5cm}
  +\zeta_1(-\nu-r-3)\biggr]
+6b_2b_1^2\Biggl[ \bigl(\gamma^2_{\rm E}-\frac{11}{3}\gamma_{\rm E}-\zeta_2 +\frac{38}{9}\bigr)\zeta(-\nu-r-3)
 \nonumber\\  &&\hspace{0.5cm} + 2\left(\gamma_{\rm E}-\frac{11}{6}\right)\zeta_1(-\nu-r-3)
 +\zeta_2(-\nu-r-3)\Biggr]
\nonumber \\&&
+ b_1^4\biggl[ \zeta_4(-\nu-r-3)+ 4(\gamma_{\rm E}-1) \, \zeta_3(-\nu-r-3) +  6\Bigl((\gamma_{\rm E}-1)^2-   \zeta_2\Bigr)  \,  \zeta_2(-\nu-r-3) \nonumber \\
  &&\hspace{1cm}
  + 4\Bigl((\gamma_{\rm E}-1)\Bigl((\gamma_{\rm E}-1)^2-3\zeta_2\Bigr)+2 \zeta_3\Bigr)
  \, \zeta_1(-\nu-r-3) \nonumber \\
  &&\hspace{1cm}+ \Bigl[(\gamma_{\rm E}-1)^2\Bigl((\gamma_{\rm E}-1)^2-6\zeta_2\Bigr)+8(\gamma_{\rm E}-1) \zeta_3 + 3\zeta_2^2-6\zeta_4\Bigr]\zeta(-\nu-r-3)  \biggr]
\label{tRi}
\eea
and
\be
\zeta_n(\nu)=  {\rm Li}_{\nu,n}(z=1)=\sum_{m=1}^{\infty} \, \frac{\ln^n m}{m^{\nu}}
\, .
   \label{zetaknu}
\ee

Strictly speaking,
the series representation (\ref{zetaknu}) for the functions $\zeta_n(-m-\nu-r -k)$
is not a good definition for large $r$ values
and we can replace them by $\zeta_n(m+\nu+r+k)$
using the result (\ref{ze_nu.1}). However, the results are long and presented in Appendix D.

\subsection{The case $\nu=1$
}

For the case $\nu=1$ we immediately have
\bea
&&A^{(i+1)}_{\rm MA,i}(Q^2) = A^{(1)}_{\rm MA,i}(Q^2) + \sum_{m=1}^{i}  \, \tilde{\delta}^{(m+1)}_{\rm MA,\nu=1,i}(Q^2) \, ,\label{tdAmanNew.1a}\\
&&\delta^{(m+1)}_{\rm MA,i}(L) \equiv \tilde{\delta}^{(m+1)}_{\rm MA,\nu=1,i}(Q^2) = 
\frac{(-1)}{m!} \, \sum_{r=0}^{\infty} \tilde{R}_m(1+r) \, \frac{(-L_i)^r}{r!} \, ,
\label{tdAmanNew.1}
\eea
where $A^{(1)}_{\rm MA,i}(Q^2)$ is given above in (\ref{tAMA1New}) (with the replacement ($L \to L_i$)  and the coefficients
$\tilde{R}_m(1+r)$ can be found in (\ref{tRi}) when $\nu=1$.


The results (\ref{tdAmanNew.1}) can be expressed in terms of the functions $\zeta_n(m+\nu+r+k)$. Using the results in Appendix D and taking the even part
($r=2m$) and the odd part ($r=2s-1$) (see equation (\ref{tdAmanNew})), we have 
\bea
&&\delta^{(2)}_{{\rm MA},k}(Q^2)= \frac{2}{(2\pi)^2} \, \Biggl[\sum_{m=0}^{\infty} (2m+1)(-1)^m Q_{1a}(2m+2)\hat{L}_k^{2m} -\pi \sum_{s=1}^{\infty} s(-1)^s Q_{1b}(2s+1)\hat{L}_k^{2s-1}
  \Biggr], \nonumber \\
&&\delta^{(3)}_{{\rm MA},k}(Q^2)= -\frac{1}{(2\pi)^3} \, \Biggl[\pi \sum_{m=0}^{\infty} (2m+1)(m+1)(-1)^m Q_{2b}(2m+3)\hat{L}_k^{2m} \nonumber \\&&\hspace{1cm}
  +2 \sum_{s=1}^{\infty} s(2s+1)(-1)^s Q_{2a}(2s+2)\hat{L}_k^{2s-1}
  \Biggr], \nonumber \\
&&\delta^{(4)}_{{\rm MA},k}(Q^2)= -\frac{2}{3(2\pi)^4} \, \Biggl[\sum_{m=0}^{\infty} (2m+1)(m+1)(2m+3)(-1)^m Q_{3a}(2m+4)\hat{L}_k^{2m} \nonumber \\&&\hspace{1cm}
  -\pi \sum_{s=1}^{\infty} s(2s+1)(s+1)(-1)^s Q_{3b}(2s+3)\hat{L}_k^{2s-1}
  \Biggr], \nonumber \\
&&\delta^{(5)}_{{\rm MA},k}(Q^2)= \frac{1}{6(2\pi)^5} \, \Biggl[\pi \sum_{m=0}^{\infty} (2m+1)(m+1)(2m+3)(m+2)(-1)^m Q_{4b}(2m+5)\hat{L}_k^{2m} \nonumber \\&&\hspace{1cm}
  +2 \sum_{s=1}^{\infty} s(2s+1)(s+1)(2s+3)(-1)^s Q_{4a}(2s+4)\hat{L}_k^{2s-1}
  \Biggr],
\label{tdAmanNew1}
\eea
where
\be
\hat{L}_k=\frac{L_k}{2\pi}
\label{hL}
\ee
and the function $Q_{ma}$ and $Q_{mb}$ are given in Appendix D.\\

At the point $L_k=0$, i.e. $Q^2=\Lambda_k^2$, we have
\bea
&&A^{(1)}_{\rm MA}= \frac{1}{2},~~~
\delta^{(2)}_{s}= \frac{2}{(2\pi)^2} \, Q_{1a}(2)= -\frac{b_1}{2\pi^2} \, \Bigl(\zeta_1(2)+l\zeta(2)\Bigr),\nonumber \\
&&\delta^{(3)}_{s}= -\frac{\pi}{(2\pi)^3} \, Q_{2b}(3)=  \frac{b_1^2}{4\pi^2} \, \Bigl(\zeta_1(3)+(2l-1)\zeta(3)\Bigr),\nonumber \\
&&\delta^{(4)}_{s}= -\frac{2}{(2\pi)^4} \, Q_{3a}(4)= \frac{1}{8\pi^4} \, \Biggl[3b_1b_2 \Bigl(\zeta_1(4)+l\zeta(4)\Bigr) - \frac{b_3}{2} \zeta(4)\nonumber \\
  &&+b_1^3\biggl\{\zeta_3(4)
  +3\left(l-\frac{5}{6}\right)\zeta_2(4) +\left(3l^2-5l-2-\frac{3\pi^2}{4}\right)\zeta_1(4) \nonumber \\
 &&+  \left(l^3-\frac{5}{2}l^2-2l+\frac{1}{2}+\frac{3\pi^2}{4}\right)\zeta(4)\biggr\}\Biggr],\nonumber \\
&&\delta^{(5)}_{s}= \frac{\pi}{(2\pi)^5} \, Q_{1b}(5)= -\frac{b_1}{8\pi^4} \, \Biggl[3b_1b_2 \Bigl(\zeta_1(5)+(l-\frac{1}{4})\zeta(5)\Bigr)  -
  \frac{b_3}{2} \zeta(5)\nonumber \\
 && +b_1^3\biggl\{
  \zeta_3(5)
  +3\left(l-\frac{13}{12}\right)\zeta_2(5)+3\left(l^2 -\frac{13}{6}l -\frac{1}{4}-\frac{\pi^2}{12}\right)\zeta_1(5) \nonumber \\
 && + \left(l^3-\frac{13}{12}l^2-\frac{3}{4}l+1+\frac{\pi^2}{4}\right)\zeta(5)\biggr\}\Biggr],
\label{dAmanNew2}
\eea
where $\zeta_k(\nu)$ are given in Eq. (\ref{zetaknu}) and 
\be
l=\ln(2\pi) \, .
\label{l2pi}
\ee

\section{Integral representations for minimal analytic coupling }

As already discussed in Introduction, the MA couplant $A^{(1)}_{\rm MA}(Q^2)$ is constructed as follows: 
the LO
spectral function is taken directly from perturbation theory
but the MA couplant $A^{(1)}_{\rm MA}(Q^2)$ itself was  built  using the correct integration counter.
Thus, at LO, the MA couplant $A^{(1)}_{\rm MA}(Q^2)$ obeys Eq. (\ref{disp_MA_LO}) presented in Introduction.

For the $\nu$-derivative of $A^{(1)}_{\rm MA}(Q^2)$, i.e. $\tilde{A}^{(1)}_{\rm MA,\nu}(Q^2)$, there is the following
equation \cite{GCAK}:
\be
\tilde{A}^{(1)}_{\rm MA,\nu}(Q^2)=
\frac{(-1)}{
  \Gamma(\nu)}
\int_{0}^{\infty} \ \frac{d s}{s} r^{(1)}_{\rm pt}(s)
    {\rm Li}_{1-\nu} (-sz)\,,
\label{disptAnuz} 
\ee
where
$r^{(1)}_{\rm pt}(s)$ is the LO spectral function defined in Eq. (\ref{SpeFun_LO})
and ${\rm Li}_{1-\nu} (-sz)$ is the Polylogarithmic function presented in (\ref{Linu}).

Beyond LO, Eq. (\ref{disptAnuz}) can be extended
in two ways, which will be shown in following subsections.

\subsection{Modification of spectral functions}

The first possibility to extend the result (\ref{disptAnuz}) beyond LO is related to the modification of the spectral function.
The extension is simple and the final result looks like this:
\be
\tilde{A}^{(i+1)}_{{\rm MA},\nu,k}(Q^2) =
\frac{(-1)}{
  \Gamma(\nu)}
\int_{0}^{\infty} \ \frac{d s}{s} r^{(i+1)}_{\rm pt}(s)
    {\rm Li}_{1-\nu} (-sz_k)\,,
    \label{disptAnuz.ma}
    \ee
    i.e. it is similar to (\ref{disptAnuz}) with the replacement the LO spectral function $r^{(1)}_{\rm pt}(s)$ by $i+1$-order one $r^{(i+1)}_{\rm pt}(s)$:
\be
r^{(i+1)}_{\rm pt}(s)=r^{(1)}_{\rm pt}(s)+ \sum_{m=1}^{i} \delta^{(m+1)}_{\rm r}(s) \, 
\label{rima.1}
\ee
and
(see \cite{NeSi,Nesterenko:2017wpb})
\bea
&&y=\ln s,~~~r^{(1)}_{\rm pt}(y)=\frac{1}{y^2+\pi^2} \,,~~~
\delta^{(2)}_{\rm r}(y)=-\frac{b_1}{(y^2+\pi^2)^2} \, \Bigl[2y f_1(y) +(\pi^2-y^2) f_2(y)\Bigr] \,,\nonumber \\
&&\delta^{(3)}_{\rm r}(y)=\frac{b_1^2}{(y^2+\pi^2)^3} \, \biggl[(3y^2-\pi^2) \Bigl\{\frac{b_2}{b_1^2} + f_1(y)\bigl(f_1(y)-1\bigr) -\pi^2 f_2^2(y)-1\Bigr\}
  \nonumber \\ && \hspace{2cm}
  -y(y^2-3\pi^2) f_2(y)\bigl(2f_1(y)-1\bigr)\biggr]\,, \nonumber \\
&&\delta^{(4)}_{\rm r}(y)=\frac{b_1^3}{(y^2+\pi^2)^4} \, \biggl[4y(y^2-\pi^2) \Bigl\{f_1(y)\bigl(3\pi^2f_2^2(y)-f_1^2(y)\bigr)
  +\frac{5}{2} \bigl(f_1^2(y)-\pi^2f_2^2(y)\bigr)  \nonumber \\ &&  \hspace{2cm}
  + f_1(y) \left(2- 3\frac{b_2}{b_1^2}\right) +\frac{1}{2} \left(\frac{b_3}{b_1^3}-1\right)\Bigr\}
  +\bigl[4\pi^2y^2-(y^2-\pi^2)^2\bigr] f_2(y) \nonumber \\ &&  \hspace{2cm}
  \times \Bigl\{\pi^2 f_2^2(y) -3 f_1^2(y) + 5 f_1(y) -3\frac{b_2}{b_1^2}+2 \Bigr\}\biggr] \,, \nonumber \\
&&\delta^{(5)}_{\rm r}(y)=\frac{1}{(y^2+\pi^2)^5} \, \biggl[(5y^4-10\pi^2y^2+\pi^4) \Bigl\{b_1^4 \, F_1(y) -3b_1^2b_2 \Bigl(1+f_1(y)
 \nonumber \\ &&
 \hspace{2cm}-2\bigl[f_1^2(y)-\pi^2f_2^2(y)\bigr]\Bigr)
 -b_1b_3 \left(\frac{1}{6}+2f_1(y)\right)+\frac{1}{3}\Bigl(b_4+5b_2^2\Bigr)\Bigr\}
\nonumber \\ &&
 \hspace{2cm}
 -y(y^4-10\pi^2y^2+5\pi^4)f_2(y) \Bigl\{b_1^4 \, F_2(y) +3b_1^2b_2 \bigl(4f_1(y)-1\bigr) - 2b_1b_3 
  \Bigr\}
  \biggr] \, , \label{dr5ma}
\eea
with
\be
f_1(y)=\frac{1}{2} \, \ln\bigl(y^2+\pi^2\bigr),~~f_2(y)=\frac{1}{2} -\frac{1}{\pi}\, arctan\left(\frac{y}{\pi}\right) \, .
 \label{f12}
\ee
and
\bea
&&F_1(y)=\frac{7}{6} + 4f_1(y) - \frac{3}{2} \bigl(f_1^2(y)-\pi^2f_2^2(y)\bigr) - \frac{13}{3} f_1(y) \,\bigl(f_1^2(y)-3\pi^2f_2^2(y)\bigr)\nonumber \\
&&\hspace{2cm}
+\bigl(f_1^4(y)-6\pi^2f_1^2(y)f_2^2(y)+\pi^4f_2^4(y)\bigr), \nonumber \\
&&F_2(y)=4-3f_1(y) - \frac{13}{3} \bigl(3f_1^2(y)-\pi^2f_2^2(y)\bigr) + 4f_1(y) \,\bigl(f_1^2(y)-\pi^2f_2^2(y)\bigr) \, .
 \label{Fi}
\eea

  For the couplant itself, we have
   \be
   A^{(i+1)}_{\rm MA,k}(Q^2) \equiv \tilde{A}^{(i+1)}_{{\rm MA},\nu=1,k}(Q^2) 
   =
   \int_0^{+\infty} \,  
\frac{ d s \, r^{(i+1)}_{\rm pt}(s)}{(s + t_k)}\,.
\label{dispA.ma} 
\ee

Numerical evaluations of the integrals in (\ref{dispA.ma}) can be done following to discussions in Section 4 in Ref. \cite{NeSi}.

\subsection{Modification of Polylogaritms}

Beyond LO, the results (\ref{disptAnuz}) can be extended also by using the $\hat{R}_m$ operators shown in (\ref{hR_i}).
This is the  path already used in Sections 4 and 5 to obtain other $\tilde{A}^{(i+1)}_{{\rm MA},\nu,i}(Q^2)$ results.

Here, the application of the operators $\hat{R}_m$ for Eq. (\ref{disptAnuz}) leads to
the following result:
\be
\tilde{A}^{(i+1)}_{{\rm MA},\nu,i}(Q^2) =
\int_{0}^{\infty}  \frac{d s}{s} r^{(1)}_{\rm pt}(s)
\tilde{\Delta}^{(i+1)}_{\nu,i}\,,
    \label{disptAnuz.ma1}
    \ee
where the results for $\tilde{\Delta}^{(i+1)}_{\nu,i}$ can be found in Eqs. (\ref{tAMAnu.1}), (\ref{tAMAnu.2}), (\ref{oRi}) and also in Eqs. (\ref{tAMAnu.2a})-(\ref{Pkz}).

For MA couplant itself, we have beyond LO 
\be
A^{(i+1)}_{\rm ma,i}(Q^2) \equiv \tilde{A}^{(i+1)}_{{\rm ma},\nu=1,i}(Q^2) 
=
\int_0^{+\infty} \,  
 \frac{d s}{s} r^{(1)}_{\rm pt}(s)
\tilde{\Delta}^{(i+1)}_{\nu=1,i}\, ,
\label{dispA.ma} 
\ee
where the
results for $\tilde{\Delta}^{(i+1)}_{\nu=1,i}$ are given in Eq. (\ref{tAMAnu.2a}) with $\nu=1$, i.e. 
\be
\tilde{\Delta}^{(i+1)}_{\nu=1,i}=\tilde{\Delta}^{(1)}_{1,i} +\sum_{m=1}^i \, \frac{P_{m,1}(z_i)}{m!}=\Delta^{(1)}_{1,i} +\sum_{m=1}^i \, \frac{P_{m,1}(z_i)}{m!}\,,
\label{tAMAnu.2aN}
\ee
where $\Delta^{(1)}_{1,i}={\rm Li}_0(z_i)$ and 
$P_{m,1}(z_i)$ are given in Eq. (\ref{PkZm}).\\

\subsection{Discussions}

\begin{figure}[!htb]
\centering
\includegraphics[width=0.58\textwidth]{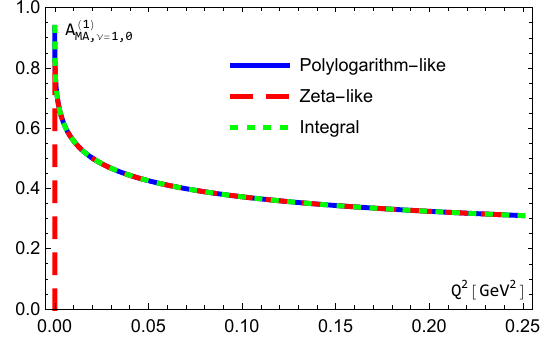}
    \caption{\label{fig:A1Leg}
      The results for $A^{(1)}_{\rm MA,\nu=1,0}(Q^2)$ with
      $\Lambda_0^{f=3}$. The Polylogarithm-like, zeta-like and integral (\ref{disptAnuz.ma}) forms
      have been used.}
\end{figure}

\begin{figure}[!htb]
\centering
\includegraphics[width=0.58\textwidth]{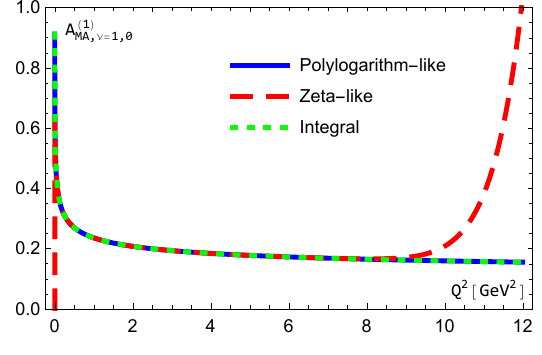}
    \caption{\label{fig:A1wide}
Same as in Fig.(\ref{fig:A1Leg}) but for larger $Q^2$ values.}
\end{figure}

\begin{figure}[!htb]
\centering
\includegraphics[width=0.58\textwidth]{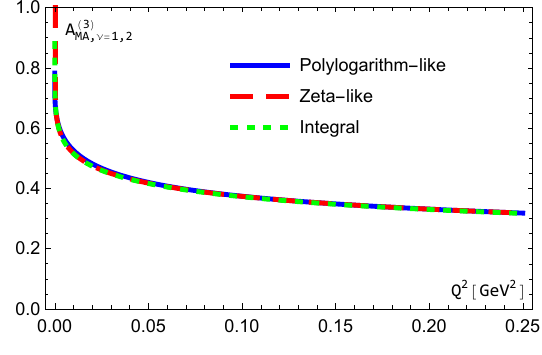}
    \caption{\label{fig:All3}
      The results for $A^{(3)}_{\rm MA,\nu=1,2}(Q^2)$ with
      $\Lambda_2^{f=3}$. The Polylogarithm-like, zeta-like and integral forms
      have been used.}
\end{figure}

\begin{figure}[!htb]
\centering
\includegraphics[width=0.58\textwidth]{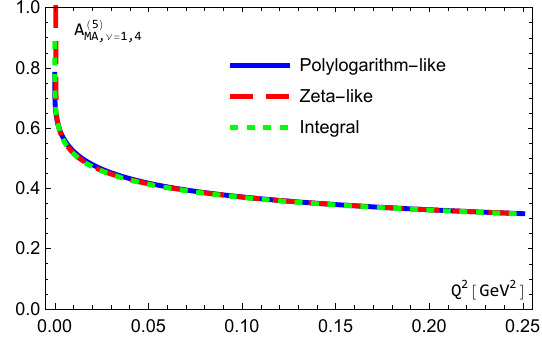}
    \caption{\label{fig:All5}
      The results for $A^{(5)}_{\rm MA,\nu=1,4}(Q^2)$ with
      $\Lambda_4^{f=3}$. The Polylogarith-like, zeta-like and integral forms
      have been used.}
\end{figure}

On Figs. \ref{fig:A1Leg}, \ref{fig:A1wide}, \ref{fig:All3} and \ref{fig:All5}  the results (\ref{tAiman.1}) (we call this ``Polylogarithm-like''),
  the results (\ref{tdAmanNew.1a})  (we call this ``Zeta-like'')
  and the results (\ref{disptAnuz.ma}) (we call them ``integral forms'') are shown. From  Figs. \ref{fig:A1Leg} and \ref{fig:A1wide} we see that
  at $0.02$ GeV$^2$ $\leq Q^2 \leq 9$ GeV$^2$ all results are very close to each other and  are indistinguishable. So, all the results are indeed the same, as they should be.

Note, however,  that the "Polylogarithm-like" results (see, for example, Eq. (\ref{tAiman.1})) consist of two parts, each of which is singular at
    the point $Q^2 = \Lambda_i^2$. The level of singularities increases with the order of the perturbation theory. So that in the ``Polylogarithmic-like''
    case the resulting curves in Figs. \ref{fig:A1Leg}, \ref{fig:A1wide}, \ref{fig:All3} and \ref{fig:All5} in $Q^2 = \Lambda_i^2$, were continuous at
    the point $Q^2 = \Lambda_i^2$,
    we should expand each part as $Q^2 \to \Lambda_i^2$ to cancel all singularities coming from both parts.

Results (\ref{tdAmanNew.1a}), (\ref{tdAmanNew.1})
are poorly applicable for $Q^2 \to 0$ and  $Q^2 \to \infty$, as they should be, because we actually use only a finite number of terms
$(r \leq 100)$ on the right-hand side of  (\ref{tdAmanNew.1a}).
However, the results of (\ref{tdAmanNew.1a})-(\ref{tdAmanNew1}) are very good for $Q^2$ intermediate  values: $0.1$ GeV$^2 < Q^2 < 10$ GeV$^2$.
Increasing  the value of $r$ leads to an expansion of the $Q^2$ range, where the results (\ref{tdAmanNew.1a}), (\ref{tdAmanNew.1}) are applicable.

Since all the results presented in Fig. \ref{fig:A1Leg}, \ref{fig:A1wide}, \ref{fig:All3} and \ref{fig:All5} are indeed the same, for
applications we can use the one that is most convenient in each particular case.

Our paper is quite long, as it contains many formulas. Thus, we will postpone the main applications of the formulas for our future publications
(see the discussion in the conclusion). Here we will consider only one application: we will study the Bjorken sum rule.
\footnote{
  The applicability of the MA approach for the Bjorken sum rule has been studied in Refs.
  \cite{Pasechnik:2009yc,Khandramai:2011zd,Pasechnik:2008th,Ayala:2017uzx,Ayala:2018ulm}.}
We will follow previous research in Refs. 
\cite{Pasechnik:2009yc,Khandramai:2011zd,Pasechnik:2008th,Ayala:2017uzx,Ayala:2018ulm,Chen:2006tw,Chen:2005tda,Kotikov:2012eq}
(see also the
Charter IV.8 in Ref. \cite{Enterria}).

\section{Bjorken sum rule}

The polarized Bjorken sum rule
is defined as the difference between proton and neutron polarized structure function
$g_1$ integrated over the whole $x$ interval
\be
\Gamma_1^{p-n}(Q^2)=\int_0^1 \, dx\, \bigl[g_1^{p}(x,Q^2)-g_1^{n}(x,Q^2)\bigr].
\label{Gpn} 
\ee

Based on the various measurements of these structure functions, the inelastic part of the above quantity, $\Gamma^{p-n}_{1,\rm inel.}(Q^2)$,
has been extracted at various values of squared momenta $Q_j^2$ (0.054 GeV$^2 \leq Q_j^2 <$ 5 GeV$^2$).

Theoretically, the quantity can be written in the Operator Product Expansion
form
\be
\Gamma_1^{p-n}(Q^2)=\left|\frac{g_A}{g_V}\right| \, \frac{1}{6} \, \bigl(1-D_{\rm BS}(Q^2)\bigr) + \sum_{i=2}^{\infty} \frac{\mu_{2i}(Q^2)}{Q^{2i-2}} \, ,
\label{Gpn.OPE} 
\ee
where $|g_A/g_V|$=1.2723 $\pm$ 0.0023 is the ratio of the nucleon axial charge, $(1-D_{BS}(Q^2))$ is the perturbation expansion for the leading-twist
contribution, and $\mu_{2i}/Q^{2i-2}$ is the higher-twist contributions.

The
twist-four term \cite{Shuryak:1981kj,Shuryak:1981pi} can be expressed at LO
\footnote{For the power-like corrections we restrict ourselves by LO approximation.}
as \cite{Chen:2006tw,Chen:2005tda}
(see discussions in Ref.  \cite{Pasechnik:2009yc}):
\be
\mu_{4}(Q^2)=\mu_{4}(Q^2_0) \, {\left[\frac{a_s(Q^2)}{a_s(Q^2_0)}\right]}^{d_4},~~ d_4=\frac{32}{9\beta_0}\, ,
\label{mu4Q2} 
\ee
which is modified in the MA case as (see \cite{Pasechnik:2009yc})
\be
\mu_{4}(Q^2)=\mu_{4}(Q^2_0) \, \frac{\tilde{A}^{(1)}_{{\rm MA},d_4,0}(Q^2)}{\tilde{A}^{(1)}_{{\rm MA},d_4,0}(Q^2_0)}\, .
\label{mu4Q2MA} 
\ee

Since we will consider very low $Q^2$ values, the above representation (\ref{Gpn.OPE}) of the
higher-twist contributions are not so convenient and it is better to use so-called its ``massive'' counter-part (following to Ref. \cite{Teryaev:2013qba}):
\be
\Gamma_1^{p-n}(Q^2)=\left|\frac{g_A}{g_V}\right| \, \frac{1}{6} \, \bigl(1-D_{\rm BS}(Q^2)\bigr) +\frac{\tilde{\mu}_4}{Q^{2}+M^2} \, ,
\label{Gpn.mOPE} 
\ee
where the values of $\tilde{\mu}$ and $M^2$ has been fitted in the papers \cite{Ayala:2017uzx,Ayala:2018ulm}
in the different types of
models for analytic QCD.

The perturbative part has the following form
\be
D_{\rm BS}(Q^2)=\frac{4}{\beta_0} \, a_s\left(1+d_1a_s+d_2a_s^2+d_3a^3_s\right)=
\frac{4}{\beta_0} \, \left(\tilde{a}_{1}+\tilde{d}_1\tilde{a}_2+\tilde{d}_2\tilde{a}_3+
\tilde{d}_3\tilde{a}_4\right),
\label{DBS} 
\ee
where
\be
\tilde{d}_1=d_1,~~\tilde{d}_2=d_2-b_1d_1,~~\tilde{d}_3=d_3-\frac{5}{2}b_1d_2+\bigl(\frac{5}{2}b^2_1-b_2\bigr)\,d_1\,.
\label{tdi} 
\ee
Eq. (\ref{DBS}) contains two expansions: one in powers of the strong coupling $a_s$ and the other in fractional derivatives $\tilde{a}_i$
  defined in Section III.  The coefficients $\tilde{d}_i$ in front of the couplants $\tilde{a}_i$ are obtained from the coefficients $d_i$ by expanding the powers of usual strong
  couplant $a_s$ through the ones $\tilde{a}_i$ (see Appendix B).

For $f=3$ case, we have
\be
\tilde{d}_1=1.59,~~\tilde{d}_2=2.51,~~\tilde{d}_3=10.58 \, .
\label{td123} 
\ee

In the MA model, the perturbative part in the first ($k=1$), the second ($k=2$), third ($k=3$) and the forth ($k=4$) orders of perturbation theory has
the following form
\be
D_{\rm{MA,BS}}(Q^2) =\frac{4}{\beta_0} \, \Bigl(A^{(k)}_{\rm MA,k-1}
+ \sum_{m=2}^{k} \, \tilde{d}_{m-1} \, \tilde{A}^{(k)}_{\rm MA,\nu=m, k-1} \Bigr)\,.
\label{DBS.ma} 
\ee

Moreover, from \cite{Ayala:2018ulm} it is possible tom see that in (\ref{Gpn.mOPE})
\be
M^2=0.439,~~\tilde{\mu}_4=-0.082 \, .
\label{M,mu} 
\ee

\subsection{Discussions}

\begin{figure}[!htb]
\centering
\includegraphics[width=0.58\textwidth]{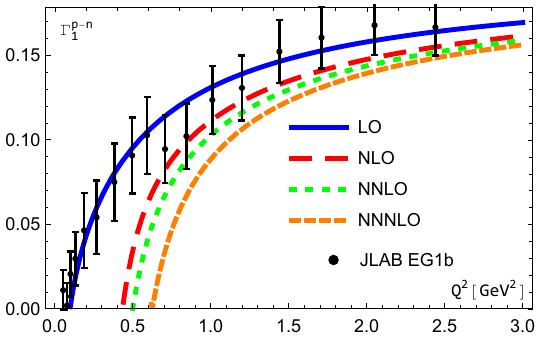}
    \caption{\label{fig:PTHT}
      The results for $\Gamma_1^{p-n}(Q^2)$ in the first four orders of perturbation theory
with the ``massive'' twist-four term (\ref{Gpn.mOPE}).}
\end{figure}

\begin{figure}[!htb]
\centering
\includegraphics[width=0.58\textwidth]{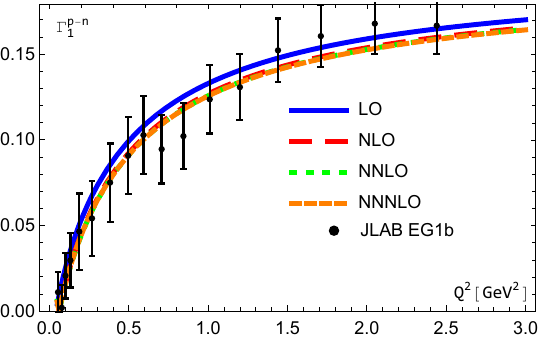}
    \caption{\label{fig:APTHT}
Same as in Fig. (\ref{fig:PTHT}) but in analytic theory.
    }
\end{figure}

The results of calculations are shown in Figs. \ref{fig:PTHT} and \ref{fig:APTHT}.
Here we use the $Q^2$-independent $M$ and $\tilde{\mu}_4$ values taken from (\ref{M,mu}) and the twist-two parts shown in Eqs. (\ref{DBS}) and (\ref{DBS.ma}) for the cases
of usual perturbation theory and APT, respectively.

As can be seen in Fig. \ref{fig:PTHT},
results obtained using usual couplants are good only at LO and deteriorate as the
order of perturbation theory increases.
The good agreement at LO is due to the use of $\Lambda_{\rm LO}$, which is small (see (\ref{Lambdas})), and therefore the investigated range of
$Q^2$ is higher than $\Lambda^2_{\rm LO}$.
Visually, the results are close to those obtained in ref. \cite{Khandramai:2011zd}, where the twist-four form (\ref{mu4Q2}) has been used.
Thus, the usage of the ``massive'' twist-four form (\ref{Gpn.mOPE}) does not improve the results, since at $Q^2 \to \Lambda_i^2$ usual couplants become to be singular, that
leads to large and negative results for the twist-two part (\ref{DBS}). With increasing the perturbation theory order usual couplants become to be singular at larger
$Q^2$ values (see Fig. \ref{fig:as1352}) and the Bjorken sum rule tends to negative values with increasing values of $Q^2$.
Thus, the discrepancy between theory and experiment increases with the increase in the order of the perturbation theory. 

In the case of using MA couplants, our results are close to those obtained in Ref. \cite{Ayala:2018ulm}, which is not surprising, since we used the parameters (\ref{M,mu})
obtained in \cite{Ayala:2018ulm}.
Moreover, we see that the results based on different orders of perturbation theory are close to each other,
in contrast to the case of using the usual couplants.


So, we see that our results for $\Gamma_1^{p-n}(Q^2)$ in the framework of usual and MA strong couplants are very similar to ones, obtained 
\cite{Khandramai:2011zd} and \cite{Ayala:2017uzx,Ayala:2018ulm}, respectively.
In future we plan to extend our present investigations for other (non-minimal) versions of analytic couplants (see \cite{CPCCAGC,3dAQCD}
and will study for the Bjorken sum rule  $\Gamma_1^{p-n}(Q^2)$
in the framework of these versions.

\section{Conclusions}

In this paper
we have considered $1/L$-expansions of  $\nu$-derivatives of the strong couplant $a_s$ expressed as
combinations of operators $\hat{R}_m$ (\ref{hR_i}) applied to the LO couplant $a_s^{(1)}$.
Applying the same operators to the $\nu$-derivatives of the LO
MA couplant $A_{\rm MA}^{(1)}$,  we obtained four different
 representations for the $\nu$-derivatives of the MA couplants, i.e. $\tilde{A}_{\rm MA, \nu}^{(i)}$, in each $i$-order of perturbation theory:
 one form contains a combination of Polylogariths;
 the other contains an expansion of the generalized Euler $\zeta$-function, and the third  is based on dispersion integrals containing
 the LO spectral function.
We also obtained a fourth representation based on the dispersion integral containing 
the  $i$-order spectral function.
All results are presented up to the 5th order of perturbation theory, where the corresponding coefficients of the QCD $\beta$-function are well known
(see \cite{Baikov:2016tgj,Herzog:2017ohr}).

The high-order corrections are negligible in the $Q^2 \to 0$ and $Q^2 \to \infty$ asymptotics and are nonzero in  the vicinity of
the point $Q^2 =\Lambda^2$. Thus,  in fact, they are really only small corrections to the LO MA couplant $A_{\rm MA,\nu}^{(1)}(Q^2)$.
This proves the possibility of expansions of high-order couplants $A_{\rm MA,\nu}^{(i)}(Q^2)$ via the LO couplants $A_ {\rm MA,\nu} ^ {(1)}(Q^2)$,
which was done in Ref. \cite{Bakulev:2010gm}, as well as the possibility of various approximations used in
\cite{Pasechnik:2008th,Pasechnik:2009yc,Khandramai:2011zd,Kotikov:2010bm,Illa}.

  As can be clearly seen, all our results ({\it up to the 5th order of perturbation theory}) have a compact form and do not contain complicated special functions,
  such as the Lambert $W$-function \cite{Magradze:1999um}, which already appears at the two-loop order as an exact solution to the usual couplant and which was used
  to evaluate  MA couplants in \cite{Bakulev:2012sm}.

  As a example, we examined
  the Bjorken sum rule and obtained results similar to previous studies in Refs.
\cite{Pasechnik:2009yc,Khandramai:2011zd,Pasechnik:2008th,Ayala:2017uzx,Ayala:2018ulm,Chen:2006tw,Chen:2005tda,Kotikov:2012eq},
because the high order corrections are small.
The results based on the usual perturbation theory do not not agree with the experimental data at $Q^2 \leq 1.5$ GeV$^2$. APT in the minimal version leads
to a good agreement with the experimental data when we used the ``massive'' version (\ref{Gpn.mOPE}) for high-twist contributions.

In the future, we plan to apply the obtained results to study the processes of deep-inelastic scattering (DIS) at small $Q^2$ values.
One of most important applications is  fitting 
experimental data for the DIS structure functions (SFs) $F_2(x,Q^2)$ and $F_3(x,Q^2)$ (see, e.g., Refs.
\cite{PKK,Shaikhatdenov:2009xd,Kotikov:2015zda,KK2001} and
\cite{KKPS1,KPS}, respectively). This
is one of the main ways
to define $\alpha_s(M_Z)$, the strong couplant  normalization.
We plan to use (the $\nu$-derivatives of) the MA couplant $\tilde{A}_{\rm MA,\nu}^{(i)}(Q^2)$ in our approximations, which is indeed possible,
because
in the fittings we study the SF Mellin moments (following Refs.
\cite{Barker,Kri}) and reconstruct SFs themselves at the end. This differs from the more popular approaches \cite{NNLOfits} based on numerical solutions of
the Dokshitzer-Gribov-Lipatov-Altarelli-Parisi (DGLAP) equations \cite{DGLAP}.
In the case of using the \cite{Barker,Kri}  approach, the $Q^2$-dependence of the SF moments is known exactly  in analytical form (see, e.g., \cite{Buras}): it
can be expressed in terms of $\nu$-derivatives
$\tilde{A}_{\rm MA,\nu}^{(i)}(Q^2)$, where the corresponding $\nu$-variable becomes to be $n$-dependent (here $n$ is the Mellin moment number),
and the use of $\nu$-derivatives should be crucial.
Beyond LO, in order to obtain complete analytic results for Mellin moments, we will use their analytic continuation \cite{KaKo}.

Note that after resumming for large values of the Bjorken variable $x$, the corresponding twist-four corrections for the SF $F_2(x,Q^2)$
changed sign at large $x$ (see \cite{Kotikov:2022vlo}). Thus, unlike the standard analyses performed in
\cite{Kotikov:2010bm}, in this case, when twist-four corrections change the sign, part of the power-like
terms can be  absorbed by the difference between the ordinary and the MA couplants, in the same way as this was done in the corresponding studies
\cite{CIKK09,Kotikov:2012sm}, conducted 
for low $x$ values in 
the framework of the so-called doubled asymptotic scaling approach \cite{Q2evo}.

Moreover, as the next steps, we plan to include in consideration the high-order terms obtained in the case of other MA couplants
(following to Refs. \cite{Bakulev:2006ex,Bakulev:2010gm,Ayala:2018ifo,Mikhailov:2021znq}), as well as in case of non-minimal versions of analytic couplants
(following to Refs. \cite{Cvetic:2006mk,Cvetic:2006gc,Cvetic:2010di,CPCCAGC,3dAQCD}).
  For non-minimal versions of analytic couplants, integral representations (\ref{disptAnuz.ma}) and (\ref{disptAnuz.ma1}) can be used. They, in turn, show
  the importance of using $\nu$-derivatives of MA couplants. Indeed, in this case it is necessary to work with the spectral functions of the couplant 
  $a_s(Q^2)$, and not with its $\nu$-degree, the calculation of which requires a very complicated procedure (see \cite{Bakulev:2012sm}).


  \section{Acknowledgments}
  We are grateful to Gorazd Cvetic for initiating these studies and collaborating at the initial stage, as well as to 
  Sergey
Mikhailov for information about his papers shown in Ref. \cite{Ayala:2018ifo}. We also want to thank them for reading the manuscript and helpful comments. 
We are also grateful to Alexander
Nesterenko for the information about the 5-loop spectral function $r^{(4)}_{\rm pt}(s)$ calculated in his paper \cite{Nesterenko:2017wpb}.
  This work was supported in part by the Foundation for the Advancement of Theoretical
Physics and Mathematics “BASIS”.

\appendix
\def\theequation{A\arabic{equation}}
\setcounter{equation}{0}

\section{QCD $\beta$-function}

The results for coefficients in expression (\ref{beta}) for QCD $\beta$-function are (see, e.g. \cite{Baikov:2016tgj,Herzog:2017ohr,Luthe:2017ttg})
\bea
&&\beta_0=11-\frac{2f}{3},~~\beta_1=102-\frac{38f}{3},~~\beta_2=\frac{2857}{2}-\frac{5033f}{18}+\frac{325f^2}{54},~~\nonumber \\
&&\beta_3=\frac{149753}{6}+3564\zeta_3 - \left(\frac{1078361}{162}+\frac{6508}{27}\zeta_3\right)\,f + \left(\frac{50065}{162}+\frac{6472}{81}\zeta_3\right)\,f^2
+ \frac{1093f^3}{729},~~\nonumber \\
&&\beta_4=\frac{8157455}{16}+ \frac{621885}{2}\zeta_3-\frac{88209}{2}\zeta_4-288090\zeta_5 \nonumber \\
&&-\left(\frac{336460813}{1944}+ \frac{4811164}{81}\zeta_3-\frac{33935}{6}\zeta_4-\frac{1358995}{27}\zeta_5\right)\,f \nonumber \\
&&+\left(\frac{25960913}{1944}+ \frac{698531}{81}\zeta_3-\frac{10526}{9}\zeta_4-\frac{381760}{81}\zeta_5\right)\,f^2  \nonumber \\
&&-\left(\frac{630559}{5832}+ \frac{48722}{243}\zeta_3-\frac{1618}{27}\zeta_4-\frac{460}{9}\zeta_5\right)\,f^3
+\left(\frac{1205}{2916}- \frac{152}{81}\zeta_3\right)\,f^4 
\label{betai}
\eea

\def\theequation{B\arabic{equation}}
\setcounter{equation}{0}

\section{Details of evaluation of the fractional derivatives}

Taking the results
(\ref{as}) of the couplant $a_s(Q^2)$ we have the following results for the $1/L$-expansion of
its $\nu$-powers:
\be
\left(a^{(1)}_{s,0}(Q^2) \right)^{\nu} = \frac{1}{L^{\nu}_0},~~
\left(a^{(i+1)}_{s,i}(Q^2) \right)^{\nu} = 
\left(a^{(1)}_{s,i}(Q^2) \right)^{\nu} + \sum_{m=2}^i \, \delta^{(m)}_{\nu,i}(Q^2)
\,,~~(i=0,1,2,...)
\label{as_nu}
\ee
where $L_k$ is defined in Eq. (\ref{L}) and
\bea
&&\delta^{(2)}_{nu,k}(Q^2) = - \frac{b_1\nu \ln L_k}{L_k^{\nu+1}} ,~~
\delta^{(3)}_{\nu,k}(Q^2) =  \frac{1}{L_k^{\nu+2}} \, \left[b_1^2\left(\frac{\nu+1}{2}\ln^2 L_k-\ln L_k-1\right)+b_2\right]\, , \nonumber \\
&&\delta^{(4)}_{s,k}(Q^2) =  \frac{1}{L_k^{\nu+3}} \, \Biggl[b_1^3\left(-\frac{(\nu+1)(\nu+2)}{6}\ln^3 L_k+\frac{2\nu+3}{2} \, \,\ln^2 L_k+(\nu+1) \,\ln L_k
  -\frac{1}{2}\right) \nonumber \\
 && \hspace{2cm} -3b_1b_2\ln L_k +\frac{b_3}{2}\Biggr]\, , \nonumber \\
&&\delta^{(5)}_{\nu,k}(Q^2) = \frac{1}{L_k^{\nu+4}} \, \Biggl[b_1^4\Biggl(\frac{(\nu+1)(\nu+2)(\nu+3)}{24}\ln^4 L_k- \frac{1}{2}\left(\nu^2+4\nu+\frac{11}{3}\right)\,\ln^3 L_k
\nonumber \\&& \hspace{2cm} -\frac{\nu(\nu+2)}{2} \,\ln^2 L_k+
     \frac{3\nu+5}{2} \,\ln L_k+ \frac{3\nu+4}{6}\Biggr)
    \nonumber \\ &&  +b^2_1b_2\left(\frac{(\nu+2)(\nu+3)}{2} \,\ln^2 L_k-(\nu+2) \,(\ln L_k+1)\right)
- \frac{b_1b_3}{2}\,\left((\nu+3)\ln L_k+\frac{1}{3}\right)\nonumber \\ &&\hspace{2cm}
+\frac{b_2^2}{2} \, \left(\nu+\frac{7}{3}\right)+\frac{b_4}{3}\Biggr] \, ,
\label{ds_nu}
\eea

which is consistent with the expansions made in  
Refs. \cite{BMS1,Bakulev:2006ex}.

The $(\nu-1)$-derivative $\tilde{a}_{\nu}(Q^2)$
is related with the $\nu+l$ $(l=0,1,2,...)$ powers as follows
\be
\tilde{a}_{\nu}(Q^2)=  a_s^{\nu}(Q^2) + k_1(\nu) a_s^{\nu+1}(Q^2) + k_2(\nu) a_s^{\nu+2}(Q^2)  + k_3(\nu) a_s^{\nu+3}(Q^2)+ k_4(\nu) a_s^{\nu+4}(Q^2)+ O(a_s^{\nu+5}) \, ,
\label{tanu}
\ee
where (see \cite{GCAK})
\bea
&&k_1(\nu)=\nu b_1 \, B_1(\nu),~~
k_2(\nu)= \nu(\nu+1) \, \left(b_2 \,  B_2(\nu) + \frac{b_1^2}{2} \, B_{1,1}(\nu) \right),~~
\nonumber \\
&&k_3(\nu)=\frac{\nu(\nu+1)(\nu+2)}{2} \left(b_3 \, B_3(\nu)+b_1b_2 \, B_{1,2}(\nu)+\frac{b_1^3}{3} \, B_{1,1,1}(\nu)\right) ,~~
\nonumber \\
&&k_4(\nu)=\frac{\nu(\nu+1)(\nu+2)(\nu+3)}{6} \Biggl(b_4 \, B_4(\nu)+
b_2^2 \, B_{2,2}(\nu)+\frac{b_1b_3}{2} \, B_{1,3}(\nu)
\nonumber \\ && \hspace{2cm}
+\frac{b^2_1b_3}{2} \, B_{1,1,2}(\nu)
+\frac{b_1^3}{4} \, B_{1,1,1,1}(\nu)\Biggr) ,~~
\label{ki}
\eea
with
\bea
&&B_1(\nu)=S_1(\nu)-1,~~B_2(\nu)=\frac{\nu-1}{2(\nu+1)},~~B_{1,1}(\nu)= Z_2(\nu+1)-2S_1(\nu)+1,~~
\nonumber \\
&&B_3(\nu)=\frac{1}{6}-\frac{1}{(\nu+1)(\nu+2)},~~B_{1,2}(\nu)=\frac{\nu}{\nu+2} \, S_1(\nu+1)+\frac{2}{\nu+1} +\frac{1}{\nu+2}-\frac{11}{6},\nonumber \\
&&B_{1,1,1}(\nu)=Z_3(\nu+2)-3Z_2(\nu+1)+3S_1(\nu)-1,~~ \nonumber \\
&&B_4(\nu)=\frac{1}{12}-\frac{2}{(\nu+1)(\nu+2)(\nu+3)},~~B_{2,2}(\nu)=\frac{5}{12} +\frac{1}{\nu+1} +\frac{1}{\nu+2}-\frac{5}{\nu+3},~~\nonumber \\
&&B_{1,3}(\nu)=\left(1-\frac{6}{(\nu+2)(\nu+3)}\right) \, S_1(\nu+1)+\frac{4}{\nu+1} -\frac{1}{\nu+2}-\frac{1}{\nu+3}-\frac{13}{6},\nonumber \\
&&B_{1,1,2}(\nu)=\frac{3(\nu+1)}{\nu+3} \, Z_2(\nu+2) +\left(\frac{12}{\nu+2} +\frac{6}{\nu+3}-11\right)\, S_1(\nu+1)
-\frac{6}{\nu+1}
\nonumber \\ && \hspace{2cm}
-\frac{5}{\nu+2}-\frac{11}{\nu+3}
+\frac{38}{3},\nonumber \\
&&B_{1,1,1,1}(\nu)=Z_4(\nu+3)-4Z_3(\nu+2)+6Z_2(\nu+1)-4S_1(\nu)+1,~~
\label{Bi},
\eea
and
\bea
&&Z_4(\nu)=S_1^4(\nu)-6S_1^2(\nu)S_2(\nu)+3S_2^2(\nu) +8S_1(\nu)S_3(\nu)-6S_4(\nu),\nonumber \\
&&Z_3(\nu)=S_1^3(\nu)-3S_2(\nu)S_1(\nu)+2S_3(\nu),~~
Z_2(\nu)=S_1^2(\nu)-S_2(\nu),\nonumber \\
&&Z_1(\nu)\equiv S_1(\nu)=\Psi(1+\nu)+\gamma_{\rm E},~~S_2(\nu)=\zeta_2-\Psi'(1+\nu),\nonumber \\
&&S_3(\nu)=\zeta_3+\frac{1}{2} \, \Psi''(1+\nu),~~S_4(\nu)=\zeta_4-\frac{1}{6} \, \Psi'''(1+\nu),~~
\label{si}
\eea
with Euler constant $\gamma_{\rm E}$ and Euler functions $\zeta_2$, $\zeta_3$ and $\zeta_4$. The expression for $Z_k(\nu)$ with arbitrary $k$ can be found in
\cite{GCAK}.

After some calculations, we have
\be
\tilde{a}^{(1)}_{\nu,0}(Q^2) =\frac{1}{L_0^{\nu}},~~
\tilde{a}^{(i+1)}_{\nu,i}(Q^2) =
\tilde{a}^{(1)}_{\nu,i}(Q^2)+\sum_{m=1}^i C_m^{\nu+m} \,\tilde{\delta}^{(m+1)}_{\nu,i}(Q^2) \, ,
\label{tanu.1b}
\ee
where
\be
\tilde{\delta}^{(m+1)}_{\nu,k}(Q^2) = R_{m,k} \, \frac{1}{L_k^{\nu+m}}
\label{tdelta_mk}
\ee
and
\bea
&&R_{1,k}=b_1 \Bigl[\hat{Z}_1(\nu)
-\ln L_k\Bigr],~~R_{2,k}=b_2 + b_1^2 \Bigl[\ln^2 L_k -2 \hat{Z}_1(\nu+1)\ln L_k + \hat{Z}_2(\nu+1 )\Bigr], \nonumber\\
&&R_{3,k}=\frac{b_3}{2} + 3b_2b_1\Bigl[Z_1(\nu+2)-\frac{11}{6}-\ln L_k\Bigr]\nonumber \\
&&+ b_1^3 \Bigl[-\ln^3 L_k +3\hat{Z}_1(\nu+2) \ln^2 L_k -3 \hat{Z}_2(\nu+2)\ln L_k + \hat{Z}_3(\nu+2 )\Bigr], \nonumber\\
&&R_{4,k}=\frac{1}{3}\,\bigl(b_4+5b_2^2\bigr) + 2b_3b_1\Bigl[Z_1(\nu+3)-\frac{13}{6}-\ln L_k\Bigr]\nonumber \\
&&+ 6b_1^2b_2 \Bigl[\ln^2 L_k -2\left(Z_1(\nu+3)-\frac{11}{6}\right) \ln L_k + Z_2(\nu+3) -\frac{11}{3} \, Z_1(\nu+3)+\frac{38}{9}
    \Bigr]\nonumber\\
&&+ b_1^4 \Bigl[\ln^4 L_k-4\hat{Z}_1(\nu+3) \ln^3 L_k +6\hat{Z}_2(\nu+3) \ln^2 L_k -4 \hat{Z}_3(\nu+3)\ln L_k + \hat{Z}_4(\nu+3 )\Bigr]\,
\label{R_i}
\eea
and $C_m^{\nu+m}$ is given in Eq. (\ref{tdmp1N}) and 
\be
\hat{Z}_k(\nu)=\sum_{m=0}^{k} \,  \frac{(-1)^m \,k!}{m!(k-m)!} \,Z_{k-m}(\nu),~~ Z_0(\nu)=1 \,,
\label{Z_i}
\ee
with
\bea
&&\hat{Z}_1(\nu)=Z_1(\nu)-1,~~\hat{Z}_2(\nu)=Z_2(\nu)-2Z_1(\nu)+1,~~\hat{Z}_3(\nu)=Z_3(\nu)-3Z_2(\nu)+3Z_1(\nu)-1, \nonumber\\
&&\hat{Z}_4(\nu)=Z_4(\nu)-4Z_3(\nu)+6Z_2(\nu)-4Z_1(\nu)+1
\,,
\label{Z_123}
\eea
where $Z_i(\nu)$ are defined in Eq. (\ref{si}).

It is convenient to introduce 
the operators $\hat{R}_i$ (\ref{tdmp1N}), which can be obtained as $\hat{R}_i=R_{i,k}\bigl(\ln L_k \to -d/(d\nu)\bigr)$. In this case, we proceed to the results
   (\ref{tdmp1N}) and (\ref{hR_i}) of the main text.

\def\theequation{C\arabic{equation}}
\setcounter{equation}{0}
\section{Another form for the differences $\tilde{\Delta}^{(i+1)}_{\nu,i}$}

We would like to note that the results (\ref{tAiman}) can be rewritten in the following way
\be
\tilde{\Delta}^{(i+1)}_{\nu,i}=\tilde{\Delta}^{(1)}_{\nu,i}
+\sum_{m=1}^i \, \frac{P_{m,\nu}(z_i)}{m! \Gamma(\nu)}
\, ,
\label{tAMAnu.2a}
\ee
where
\be
P_{m,\nu}(z_i)= \overline{R}_m(z_i) \, {\rm Li}_{-\nu-m+1}(z_i)
\label{PkZm}
\ee
and, thus,
\bea
&&P_{1,\nu}(z)=b_1\Bigl[\Bigl(\gamma_{\rm E}-1\Bigr){\rm Li}_{-\nu}(z)+{\rm Li}_{-\nu,1}(z)\Bigr], \nonumber \\
&&P_{2,\nu}(z)=b_2 \,{\rm Li}_{-\nu-1}(z) + b_1^2\Bigl[{\rm Li}_{-\nu-1,2}(z) + 2(\gamma_{\rm E}-1){\rm Li}_{-\nu-1,1}(z) \nonumber \\
&&+  \Bigl((\gamma_{\rm E}-1)^2-
    \zeta_2\Bigr) \, {\rm Li}_{-\nu-1}(z) \Bigr], \nonumber \\
&&P_{3,\nu}(z)=\frac{b_3}{2} {\rm Li}_{-\nu-2}(z) +3b_2b_1\left[ \Bigl(\gamma_{\rm E}-\frac{11}{6}\Bigr){\rm Li}_{-\nu-2}(z) +{\rm Li}_{-\nu-2,1}(z)\right] \nonumber \\
&&+ b_1^3\biggl[{\rm Li}_{-\nu-2,3}(z) + 3(\gamma_{\rm E}-1) \, {\rm Li}_{-\nu-2,2}(z)
  +  3\Bigl((\gamma_{\rm E}-1)^2-
  \zeta_2\Bigr)  \, {\rm Li}_{-\nu-2,1}(z)
\nonumber \\&&
  + \Bigl((\gamma_{\rm E}-1)\Bigl((\gamma_{\rm E}-1)^2-3\zeta_2\Bigr)+2 \zeta_3\Bigr) {\rm Li}_{-\nu-2}(z) \biggr]
, \nonumber \\
&&P_{4,\nu}(z)=\frac{1}{3}\left(b_4+5b_2^2\right) {\rm Li}_{-\nu-3}(z)+2b_3b_1\left[ \Bigl(\gamma_{\rm E}-\frac{13}{6}\Bigr){\rm Li}_{-\nu-3}(z) +{\rm Li}_{-\nu-3,1}(z)\right]
\nonumber \\&&
+6b_2b_1^2\left[ \Bigl(\gamma^2_{\rm E}-\frac{11}{3}\gamma_{\rm E}-\zeta_2 +\frac{38}{9}\Bigr){\rm Li}_{-\nu-3}(z)
  + 2\left(\gamma_{\rm E}-\frac{11}{6}\right){\rm Li}_{-\nu-3,1}(z)+{\rm Li}_{-\nu-3,2}(z)\right]
\nonumber \\&&
+ b_1^4\biggl[{\rm Li}_{-\nu-3,4}(z) + 4(\gamma_{\rm E}-1) \, {\rm Li}_{-\nu-3,3}(z) +  6\Bigl((\gamma_{\rm E}-1)^2-   \zeta_2\Bigr)  \, {\rm Li}_{-\nu-3,2}(z)  \nonumber \\
  &&
  + 4\Bigl((\gamma_{\rm E}-1)\Bigl((\gamma_{\rm E}-1)^2-3\zeta_2\Bigr)+2 \zeta_3\Bigr)
  \, {\rm Li}_{-\nu-3,1}(z) \nonumber \\
  &&
  + \Bigl\{(\gamma_{\rm E}-1)^2\Bigl((\gamma_{\rm E}-1)^2-6\zeta_2\Bigr)+8(\gamma_{\rm E}-1) \zeta_3 + 3\zeta_2^2-6\zeta_4 \Bigr\} {\rm Li}_{-\nu-3}(z) \biggr]\,,
\label{Pkz}
\eea
where ${\rm Li}_{\nu,k}(z)$ is defined in Eq. (\ref{Mnuk}).

\def\theequation{D\arabic{equation}}
\setcounter{equation}{0}
\section{Alternative
  form for the couplants $\tilde{A}^{(i+1)}_{\rm MA,\nu,i}(Q^2)$}

Using the series representation (\ref{zetaknu}),
 the functions $\zeta(n,-\nu-r -k)$ in (\ref{tRi}) are not so good defined at large $r$ values 
 and by $\zeta(n,\nu+r+k)$
and we will replace them the using
the result (\ref{ze_nu.1})
as
\be
\zeta(\nu-r)= -\frac{\Gamma(\nu+r+1)}{\pi(2\pi)^{\nu+r}}\,\tilde{\zeta}(\nu+r+1),~~\tilde{\zeta}(\nu+r+1)=\sin\left[\frac{\pi}{2}(\nu+r)\right] \, \zeta(\nu+r+1) \, .
  \label{ze_nu.1N}
\ee

After come calculations we have
\be
\tilde{\delta}^{(m+1)}_{{\rm MA},\nu,k}(Q^2)=
\frac{1}{\Gamma(\nu+m)} \, \sum_{r=0}^{\infty} \frac{\Gamma(\nu+r+m)}{\pi(2\pi)^{\nu+r+m-1}} \, Q_m(\nu+r+m) 
\, \frac{(-L_k)^r}{r!}
\label{tdAmanNew}
\ee
where
\bea
&&Q_1(\nu+r+1)=b_1\Bigl[\tilde{Z}_1(\nu+r)\tilde{\zeta}(\nu+r+1)+\tilde{\zeta}_1(\nu+r+1)\Bigr], \nonumber \\
&&Q_2(\nu+r+2)=b_2\tilde{\zeta}(\nu+r+2) + b_1^2\Bigl[\tilde{\zeta}_2(\nu+r+2) + 2\tilde{Z}_1(\nu+r+1)\tilde{\zeta}_1(\nu+r+2) \nonumber \\
  &&\hspace{0.5cm} + \tilde{Z}_1(\nu+r+1)
  \tilde{\zeta}(\nu+r+2)
  \Bigr], \nonumber \\
&&Q_3(\nu+r+3)=\frac{b_3}{2} \tilde{\zeta}(\nu+r+3) +3b_2b_1\left[ \bigl(\overline{S}_1(\nu+r+2)-\frac{11}{6}\bigr)\tilde{\zeta}(\nu+r+3)+\tilde{\zeta}_1(\nu+r+3)\right]
\nonumber\\  &&\hspace{0.5cm}   + b_1^3\biggl[\tilde{\zeta}_3(\nu+r+3) + 3 \tilde{Z}_1(\nu+r+2)\, \tilde{\zeta}_2(\nu+r+3)
  +  3\tilde{Z}_2(\nu+r+2)
  \, \tilde{\zeta}_1(\nu+r+3)
  \nonumber\\  &&\hspace{0.5cm} +
  \tilde{Z}_3(\nu+r+2) \, \tilde{\zeta}(\nu+r+3) \biggr]
, \nonumber \\
&&Q_4(\nu+r+4)=\frac{1}{3}\left(b_4+5b_2^2\right) \tilde{\zeta}(\nu+r+4)
+2b_3b_1\biggl[ \bigl(\overline{S}_1(\nu+r+3)-\frac{13}{6}\bigr) \tilde{\zeta}(\nu+r+4)
   \nonumber\\  &&\hspace{0.5cm}
  + \tilde{\zeta}_1(\nu+r+4)\biggr]
+6b_2b_1^2\Biggl[ \bigl(\overline{Z}_2(\nu+r+3)-\frac{11}{3}\overline{S}_1(\nu+r+3) +\frac{38}{9}\bigr) \tilde{\zeta}(\nu+r+3)
 \nonumber\\  &&\hspace{0.5cm} + 2\left(\overline{S}_1(\nu+r+3)-\frac{11}{6}\right) \tilde{\zeta}_1(\nu+r+4)
 + \tilde{\zeta}_2(\nu+r+4)\Biggr]
\nonumber \\&&
+ b_1^4\biggl[  \tilde{\zeta}_4(\nu+r+4)+ 4\tilde{Z}_1(\nu+r+3) \,  \tilde{\zeta}_3(\nu+r+4) +  6
  \tilde{Z}_2(\nu+r+3)\, \tilde{\zeta}_2(\nu+r+4) \nonumber \\
  &&\hspace{2cm}
  + 4\tilde{Z}_3(\nu+r+3)
  \,  \tilde{\zeta}_1(\nu+r+4)
  +\tilde{Z}_4(\nu+r+3)
   \tilde{\zeta}(\nu+r+4)  \biggr]
\label{Qi}
\eea
with (see also (\ref{si})
\bea
&&\overline{Z}_4(\nu)=\overline{S}_1^4(\nu)-6\overline{S}_1^2(\nu)S_2(\nu)+3S_2^2(\nu) +8\overline{S}_1(\nu)S_3(\nu)-6S_4(\nu),\nonumber \\
&&\overline{Z}_3(\nu)=\overline{S}_1^3(\nu)-3S_2(\nu)\overline{S}_1(\nu)+2S_3(\nu),~~
\overline{Z}_2(\nu)=\overline{S}_1^2(\nu)-S_2(\nu),\nonumber \\
&&\overline{Z}_1(\nu)\equiv \overline{S}_1(\nu)=\Psi(1+\nu)+\gamma_{\rm E} -\ln(2\pi),~~S_2(\nu)=\zeta_2-\Psi'(1+\nu),\nonumber \\
&&S_3(\nu)=\zeta_3+\frac{1}{2} \, \Psi''(1+\nu),~~S_4(\nu)=\zeta_4-\frac{1}{6} \, \Psi'''(1+\nu),~~
\label{osi}
\eea
and (see also (\ref{Z_123}))
\bea
&&\tilde{Z}_1(\nu)=\overline{Z}_1(\nu)-1,~~\tilde{Z}_2(\nu)=\overline{Z}_2(\nu)-2\overline{Z}_1(\nu)+1,~~
\tilde{Z}_3(\nu)=\overline{Z}_3(\nu)-3\overline{Z}_2(\nu)+3\overline{Z}_1(\nu)-1, \nonumber\\
&&\tilde{Z}_4(\nu)=\overline{Z}_4(\nu)-4\overline{Z}_3(\nu)+6\overline{Z}_2(\nu)-4\overline{Z}_1(\nu)+1
\, .
\label{oZ_123}
\eea

Moreover we use here
\be
\tilde{\zeta}_k(\nu)=  \frac{d^k}{(d\nu)^k} \, \tilde{\zeta}(\nu) \, .
   \label{tzetaknu}
\ee

Using the definition of $\tilde{\zeta}_k(\nu)$ in (\ref{ze_nu.1N}), we have
\bea
&&\tilde{\zeta}_1(\nu+r+1)=\sin\left[\frac{\pi}{2}(\nu+r)\right] \, \zeta_1(\nu+r+1) + \frac{\pi}{2} \,
\cos\left[\frac{\pi}{2}(\nu+r)\right] \, \zeta(\nu+r+1) \, ,\nonumber \\
&&\tilde{\zeta}_2(\nu+r+1)=\sin\left[\frac{\pi}{2}(\nu+r)\right] \, \left(\zeta_2(\nu+r+1) - \frac{\pi^2}{4}\,\zeta(\nu+r+1)\right)
\nonumber \\ && \hspace{1cm}
+ \frac{\pi}{2} \,
\cos\left[\frac{\pi}{2}(\nu+r)\right] \, \zeta_1(\nu+r+1) \, ,\nonumber \\
&&\tilde{\zeta}_3(\nu+r+1)=\sin\left[\frac{\pi}{2}(\nu+r)\right] \, \left(\zeta_3(\nu+r+1) - \frac{3\pi^2}{4}\,\zeta_1(\nu+r+1)\right)
\nonumber \\ && \hspace{1cm} + \frac{\pi}{2} \,
\cos\left[\frac{\pi}{2}(\nu+r)\right] \, \left(3\zeta_2(\nu+r+1) - \frac{\pi^2}{4}\,\zeta(\nu+r+1)\right)\, ,\nonumber \\
&&\tilde{\zeta}_4(\nu+r+1)=\sin\left[\frac{\pi}{2}(\nu+r)\right] \, \left(\zeta_4(\nu+r+1) - \frac{3\pi^2}{2}\,\zeta_2(\nu+r+1)+\frac{\pi^4}{16}\,\zeta(\nu+r+1)\right)
\nonumber \\ && \hspace{1cm} + \frac{\pi}{2} \,
\cos\left[\frac{\pi}{2}(\nu+r)\right] \, \left(4\zeta_3(\nu+r+1) - \pi^2\,\zeta_1(\nu+r+1)\right)\, ,
\label{tzeta1234}
\eea
where $\zeta_k(\nu)$ are given in Eq. (\ref{zetaknu}) of the main text.

So, we can rewrite
the results (\ref{tdAmanNew}) with
\bea
&&Q_{m}(\nu+r+m)=\sin\left[\frac{\pi}{2}(\nu+r+m-1)\right] Q_{ma}(\nu+r+m) \nonumber\\
&&+ \frac{\pi}{2} \,
\cos\left[\frac{\pi}{2}(\nu+r+m-1)\right] Q_{mb}(\nu+r+m)\, ,
\label{Qmab}
\eea
where
\bea
&&Q_{1a}(\nu+r+1)=b_1\Bigl[\tilde{Z}_1(\nu+r)\zeta(\nu+r+1)+\zeta_1(1,\nu+r+1)\Bigr],~~
\nonumber \\ &&Q_{1b}(\nu+r+1)=b_1\zeta(\nu+r+1), \nonumber \\
&&Q_{2a}(\nu+r+2)=b_2\zeta(\nu+r+2) + b_1^2\Bigl[\zeta_2(\nu+r+2) + 2\tilde{Z}_1(\nu+r+1)\zeta_1(\nu+r+2) \nonumber \\
  &&\hspace{0.5cm} + \left(\tilde{Z}_1(\nu+r+1) - \frac{\pi^2}{4}\right)
  \zeta(\nu+r+2)
  \Bigr], \nonumber \\
&&Q_{2b}(\nu+r+2)=2b_1^2\Bigl[\tilde{Z}_1(\nu+r+1)\zeta(\nu+r+2)+\zeta_1(\nu+r+2)\Bigr],~~ \nonumber \\
&&Q_{3a}(\nu+r+3)=\frac{b_3}{2} \zeta(\nu+r+3) +3b_2b_1\left[ \bigl(\overline{S}_1(\nu+r+2)-\frac{11}{6}\bigr)\zeta(\nu+r+3)+\zeta_1(\nu+r+3)\right]
\nonumber\\  &&\hspace{0.5cm}   + b_1^3\biggl[\zeta_3(\nu+r+3) + 3 \tilde{Z}_1(\nu+r+2)\, \zeta_2(\nu+r+3)
  +  3\left(\tilde{Z}_2(\nu+r+2)- \frac{\pi^2}{4}\right)
  \, \zeta_1(\nu+r+3)
  \nonumber\\  &&\hspace{0.5cm} +
  \left(\tilde{Z}_3(\nu+r+2) - \frac{3\pi^2}{4}\right)\, \zeta(\nu+r+3) \biggr]
, \nonumber \\
&&Q_{3b}(\nu+r+3)= 3b_2b_1 \zeta(\nu+r+3) + 3b_1^3\biggl[\zeta_2(\nu+r+3) + 2 \tilde{Z}_1(\nu+r+2)\, \zeta_1(\nu+r+3)\nonumber\\
  &&\hspace{0.5cm}
  +\left(\tilde{Z}_2(\nu+r+2) - \frac{\pi^2}{12}\right)
  \, \zeta(\nu+r+3) \biggr], \nonumber \\
&&Q_{4a}(\nu+r+4)=\frac{1}{3}\left(b_4+5b_2^2\right) \zeta(\nu+r+4)
+2b_3b_1\biggl[ \bigl(\overline{S}_1(\nu+r+3)-\frac{13}{6}\bigr) \zeta(\nu+r+4)
   \nonumber\\  &&\hspace{0.5cm}
  + \zeta_1(\nu+r+4)\biggr]
+6b_2b_1^2\Biggl[ \bigl(\overline{Z}_2(\nu+r+3)-\frac{11}{3}\overline{S}_1(\nu+r+3) +\frac{38}{9}- \frac{\pi^2}{4}\bigr) \zeta(\nu+r+3)
 \nonumber\\  &&\hspace{0.5cm} + 2\left(\overline{S}_1(\nu+r+3)-\frac{11}{6}\right) \zeta_1(\nu+r+4)
 + \zeta_2(\nu+r+4)\Biggr]
\nonumber \\&&
+ b_1^4\biggl[  \zeta_4(\nu+r+4)+ 4\tilde{Z}_1(\nu+r+3) \,  \zeta_3(\nu+r+4) +  6
  \left(\tilde{Z}_2(\nu+r+3)- \frac{\pi^2}{4}\right)\, \zeta_2(\nu+r+4) \nonumber \\
  &&\hspace{2cm}
  + 4\left(\tilde{Z}_3(\nu+r+3)- \frac{3\pi^2}{4}\right)
  \,  \zeta_1(\nu+r+4)
  \nonumber \\&&\hspace{2cm}
  +\left(\tilde{Z}_4(\nu+r+3)- \frac{3\pi^2}{2}\tilde{Z}_2(\nu+r+3)+ \frac{\pi^4}{16}\right)
  \zeta(\nu+r+4)  \biggr], \nonumber \\
&&Q_{4b}(\nu+r+4)=2b_3b_1\zeta(\nu+r+4)+12b_2b_1^2 \biggl[ \bigl(\overline{S}_1(\nu+r+3)-\frac{11}{6}\bigr) \zeta(\nu+r+4)
   \nonumber\\  &&\hspace{0.5cm}
   + \zeta_1(\nu+r+4)\biggr]
+ 4b_1^4\biggl[\zeta_3(\nu+r+4)
  + 3 \tilde{Z}_1(\nu+r+3)\, \zeta_2(\nu+r+4)
  \nonumber \\ &&\hspace{0.5cm}
  +  3\left(\tilde{Z}_2(\nu+r+3)- \frac{\pi^2}{12}\right)
  \, \zeta_1(\nu+r+4)
  +\left(\tilde{Z}_3(\nu+r+3) - \frac{\pi^2}{4}\right)\, \zeta(\nu+r+4) \biggr]\, .
\label{Qiab}
\eea


\end{document}